# On-demand augmentation in heat transfer of Taylor bubble flows using ferrofluids

Ram Krishna Shah[1], Sameer Khandekar[1*]

[1]Department of Mechanical Engineering, Indian Institute of Technology Kanpur, UP, India

*Corresponding Author: Tel: +91-512-259-7038, Email: samkhan@iitk.ac.in

**ABSTRACT**

The thermo-fluidic transport characteristics of ferrofluids can be influenced by the application of a magnetic field. The magnetic manipulations of ferrofluids have been useful in augmenting heat transfer, as evident from recent investigations. In the present study, we examine a novel strategy for augmenting two-phase heat transfer and show that the magnetic manipulation of non-boiling Taylor bubble flow (TBF) of ferrofluids can provide on-demand augmentation. In an earlier investigation (DOI:10.1016/j.colsurfa.2020.124589), we had shown that the characteristics of the TBF of ferrofluids could be altered through external magnetic manipulations. As transport characteristics of TBFs primarily depend on their flow morphology, it was anticipated that such alteration would affect their thermal transport characteristics, which are examined in the present work. The generation of smaller bubbles and unit-cells through magnetic manipulations decreases the void fraction of the resulting TBF. In addition, a greater number of units participated in the heat exchange process compared to larger bubble-slug systems at any time instance. Such flow modifications cause considerable augmentation (which can go up to 100%) in two-phase heat transfer. The extent of augmentation depends on the applied magnetic field and induced magnetic force, homogeneous gas fraction, liquid film thickness/void fraction and flow morphology of the resulting TBF, which are examined in the present study. The application of ferrofluids in TBFs provides multiple benefits, such as suspension of nanoparticles with better thermal properties and additional functionality of the flow manipulations through external means. The proposed application with the suggested manipulation technique provides an effective alternative for an on-demand augmentation in two-phase heat transfer in low Reynolds number flows.

# NOMENCLATURE

| | |
|---|---|
| $A$ | Cross-sectional area (m$^2$) |
| $B$ | Magnetic flux density (T) |
| $B_r$ | Remnant magnetic flux density (T) |
| $D_h$ | Hydraulic diameter (m) |
| $F$ | Magnetic body force (N/m$^3$) |
| $f$ | Frequency (Hz) |
| $g$ | Acceleration due to gravity (m/s$^2$) |
| $H$ | Magnetic field strength (A/m) |
| $J$ | Superficial velocity (m/s) |
| $L$ | Length (m) |
| $M$ | Magnetization (A/m) |
| $P$ | Pressure (Pa) |
| $Q$ | Flow rate (m$^3$/s) |
| $T$ | Temperature (K) |
| $U_{TB}$ | Taylor bubble velocity (m/s) |
| $X, Y, Z$ | Cartesian axis coordinates |
| $\mu_r$ | Relative permeability |
| $\mu_0$ | Permeability of free space (H/m) |
| $\xi$ | Distance of the magnet from the closest wall of channel (m) |
| $\rho$ | Density (kg/m$^3$) |
| $\mu$ | Fluid viscosity (Pa - s) |
| $\sigma$ | Surface tension (N/m) |
| $\beta$ | Gas volume/homogeneous void fraction |

*Subscripts*

| | |
|---|---|
| $ff$ | Ferrofluid |
| $TB$ | Taylor/gas bubble |
| $UC$ | Unit-cell |
| $LS$ | Liquid slug |

**Abbreviations**

NMF          No Magnetic Field

MF            Magnetic Field

TBF          Taylor Bubble Flow

**Non-dimensional numbers**

$$\text{Bond number (Bo)} = \frac{(\rho_{liquid} - \rho_{gas})gD_h^2}{\sigma}$$

$$\text{Capillary number (Ca)} = \frac{\mu_{ff} J_{ff}}{\sigma}$$

$$\text{Reynolds number (Re)} = \frac{\rho J D_h}{\mu}$$

# 1. Introduction

The gas-liquid Taylor bubble flows (TBFs), or capillary slug flows, are a periodic/intermittent and segmented flow pattern of two phases, a lighter (gas/vapor) dispersed phase and a denser (liquid) continuous phase [1]. They are typically a surface tension dominated flow pattern of elongated (capsular-shaped bubbles having lengths several times channel diameter) gas bubbles, separating two adjacent liquid slugs in confined channels [2, 3]. The bubbles are usually separated from the channel walls by a thin liquid film depending on the wetting characteristics of the liquid. The non-phase-change gas-liquid TBF pattern is a small subset of the larger domain of two-phase flow patterns having intricate and fascinating flow phenomena [4, 5].

When a single Taylor bubble or a train of bubbles is introduced in a steady flow of single-phase liquids in confined channels, the slip motion of bubbles through the continuous liquid phase significantly alters the hydrodynamics of single-phase flow (SPF). The formation of new interfaces and motion of bubbles disrupts the boundary layer formations, and circulating flows (flow with closed streamlines sometimes referred to as Taylor vortices) are created in the wake and interfacial regions of the adjacent liquid slugs as well as in the gas bubbles [6, 7]. The generation of circulating flows intensify radial mixing in the liquid slugs, which produces a multi-fold enhancement in heat/mass transport compared to SPFs [8-10]. Several investigations have been conducted to study the circulation phenomena and their effect on the transport characteristics in slug flow systems [11, 12]. A consolidation of results pertaining to the transport characteristics of gas-liquid and liquid-liquid TBFs has been carried out by Dia et al. [13].

Mehta et al. [14-16] conducted a series of experimental investigations to study the effect of introducing a single Taylor bubble, as well as a train of bubbles, in a steady flow of liquid on its thermal transport characteristics. Such experiments were performed to delineate the thermo-hydrodynamics of pulsating heat pipes (PHPs). It was observed that the motion of bubbles through the liquid phase disturbed the flow field, which was evident from the temporal fluctuations detected in the local wall temperatures. The disturbance was significant in the wake region of the Taylor bubbles compared to the bubble front. A considerable augmentation in heat transfer (1.2 - 2 times) was observed for the bubble train flow compared to the single-phase convective flow of liquids. It was also observed that the bubble length had a significant effect on the heat transfer characteristics and smaller bubbles were more effective compared to bubbles having a length scale equivalent to heater length. Similar observations (2 - 3 times enhancement in two-phase heat transfer over SPF) have also been made by earlier investigations as well [17-19]. Vyas et al. [20] carried out an analysis of TBFs from the perspective of a single unit-cell (consisting of a gas bubble and a liquid slug). Formation of recirculating flows in the oscillating liquid plugs were observed in their investigation at the capillary length scale [20]. A comprehensive review of the flow and transport characteristics of TBFs is also presented in the dissertation works of Mehta [15] and Vyas [21]. The TBF pattern is observed in many applications such as monolith micro-reactors, two-phase mini/micro heat/mass exchangers, PHPs, fuel cells, blood flow, microfluidics and Lab-on-chip devices, to name a few [1, 2, 7, 22].

Historically, bubble pump-meters were used to measure the flow rates of liquids. In such a measurement system, the velocities of the bubble indicators were measured, and it was assumed that the liquid flow had the same velocity as the bubble. Fairbrother and Stubbs [23] showed that bubbles move somewhat faster than average liquid velocity because of the existence of a liquid film near the walls. It was shown that the slip velocity of bubbles depends on the bubble Capillary number ($Ca_{TB} = \mu_{liq} U_{TB} / \sigma$) of the flow, and a scaling relationship was also proposed to estimate liquid film thickness in their study. Taylor [24] extended their experimental work and showed that for Ca > 0.09, slip velocity assumed an asymptotic value. Based on experimental observations, qualitative sketches of the streamlines of the flows in bubble-slug motion were presented by him, some of which were later examined and confirmed by Cox [25] experimentally. Bretherton [26] presented his seminal work on the motion of long bubbles in capillary tubes, which provided a theoretical framework for the hydrodynamic analysis of the same. It was shown that the long cylindrical bubbles of diameters comparable to the tubes and lengths several times the tube diameter does not instantly rise in a vertical tube filled with denser liquid under the effect of gravity for Bond number (Bo $= \Delta \rho g D_h^2 / \sigma$) < 3.37. This value of the Bond number is usually considered to be the critical (Bo$_{cr}$) for the formation of a surface tension dominated regime of TBFs [2, 22]. A scaling relationship between film thickness and the Capillary number was also proposed by him. It is evident from the earlier investigations that the deposition of the liquid thin-film on channel walls played a crucial role in modifying the hydrodynamics of TBFs. Numerous studies have taken place in last few decades to investigate the hydrodynamics and transport characteristics of the gas-liquid and liquid-liquid Taylor bubble/Slug flows because of their intriguing flow physics and importance from the industrial point of view, a comprehensive review of which can be found in the articles cited herein [2, 7, 9, 10].

The transport phenomena of TBFs/Slug flows are intricate in nature, and their thermo-hydrodynamics depend on several parameters such as slug and bubble lengths (phase distribution or flow morphology), liquid film thickness/void fraction, bubble velocity, Reynolds (Re) and Capillary number (Ca) of the flow [17, 27]. Any variations in the flow morphology and film thickness/void fraction substantially affects the heat/mass transport in such flows [17]. Prothero and Burton [28] investigated thermal transport in segmented gas-liquid flows in capillary tubes to understand 'bolus flows' and the mass transport of oxygen in such flows. A heat transfer enhancement up to two times in TBFs over SPF was reported by them. Oliver and Write [29] investigated the effect of the homogeneous gas fraction ($\beta$) on heat transfer in slug flows in a circular tube ($D_h$ = 6.25 mm). A maximum enhancement of 2.5 times over SPF was reported and observed that the shorter slugs were better for heat transfer. A correlation for the prediction of Nusselt number in TBFs was proposed by them based on the extension of Gratz theory for laminar SPF. In subsequent work, Oliver and Hoon [30] studied the effect of slug length ($L_{SL}$) on heat transfer for the given gas fractions. The slug length was changed in the range of 1 to 80 times the tube diameter and observed that the two-phase Nusselt number varied as $L_{SL}^{-1/3}$ for the small range of $\beta$ = 0.24 - 0.39.

A comparison of the earlier results and correlations proposed by various authors for predicting the two-phase heat transfer coefficient is presented in an article by Dongwoo et al. [31]. He et al. [32] carried out numerical simulations to investigate heat transfer in Slug flows and proposed analytical modelling of the same based on one-dimensional unsteady heat conduction assuming adherent liquid film. Their proposed analytical model was validated with the simulation results and it was shown that the heat transfer was dependent on slug length and mixture velocity, similar to the earlier studies. Leung et al. [17, 19] and Walsh et al. [18, 33] performed comprehensive experimental investigations to examine the effect of slug length, gas void fraction, mixture velocity, Capillary and Prandtl number on two-phase heat transfer in gas-liquid TBFs in capillary tubes ($D_h$ = 1.5 - 2 mm). For a given mixture velocity and gas fraction, the heat transfer was observed to be dependent on the slug length. It was observed that the smaller slugs were better for heat transfer compared to large slugs. It was also observed that the size and time scale of circulations was a strong function of Ca, and it reduced with an increase in Ca. Both Leung [17] and Walsh [18] proposed different correlations for the prediction of two-phase heat transfer based on their experimental observations. Their correlations modified earlier proposed models by including the length of slugs and unit-cells. It was shown that the experimental observations are well predicted by the proposed correlations. A 2-3 times higher heat transfer was observed for TBFs over SPFs in majority of the cases.

It is evident from the earlier investigations that the smaller bubble-slug unit-cells are preferred for enhanced thermal transport in TBF/Slug flows. A higher number of smaller units consisting of faster-circulating slugs are formed, which significantly improves heat/mass transfer. One way to achieve such variations in TBFs/Slug flows is by changing the governing mechanism of bubble/slug formation from squeezing to transitional and shearing flow regimes when such flows are generated using T-junction or flow-focusing geometries [34]. Squeezing and shearing flow regimes are governed by the interfacial tension and shearing stress, respectively. At low Capillary numbers, Ca = $\mu_{liq} J_{liq}/\sigma$ < $10^{-2}$, bubbles are formed due to the build-up of squeezing pressure upstream to the interface [34]. As Ca is increased (> $10^{-2}$), shearing stress dominates the bubble formation process. Such variations in the flow regime can be achieved by increasing the flow rate of the continuous (liquid) phase, which would require additional pumping power. A supplementary active (laser, electric, magnetic or acoustic field-based control/manipulation) or passive technique (local geometrical modifications, micro-valves/other mechanical techniques) can also be applied to alter the flow hydrodynamics or fluid/interfacial properties locally, e.g., laser induced heating of the interface to vary interfacial tension locally [35, 36]. Additionally, the use of another immiscible liquid as the dispersed-phase, and nanofluids as the continuous/ dispersed-phase, has become an attractive area of research because of their superior properties over their counterparts [10, 37, 38].

The use of magnetic nanofluids or ferrofluids as the continuous phase provides an energy-efficient quasi-passive way to manipulate the bubble formation mechanism through external means without changing the mean flow rates (or Ca) of the respective phases [39]. Ferrofluids are engineered colloidal suspensions of magnetic nanoparticles or MNPs (Iron-oxide nanoparticles (IONPs) are commonly used in the synthesis) of size ~ 10 nm in a non-magnetic liquid medium (e.g. water, oils) [40]. Ferrofluids exhibit magnetic properties in addition to their improved thermo-physical properties due to the suspension of MNPs. The magnetic properties of ferrofluids can be activated by applying a magnetic field (MF), which provides an additional means to control/manipulate their flow [41, 42]. Ferrofluids have shown remarkable potential for heat transfer augmentation in single-phase laminar convective flows in externally applied MFs [43-45]. The application of an MF near the flow of ferrofluids introduces local flow modification, which improves mixing in the fluid [46]. The magnetic manipulations of ferrofluids have also been proven to be useful in many microfluidic, lab-on-chip and biomedical based applications such as sorting/capturing of particles/molecules, micro-scale pumps and valves etc. [47, 48]. In a recent study, Gui et al. [49] explored the application of ferrofluids in slug flow heat transfer in a micro-channel. A minor enhancement in two-phase heat transfer for ferrofluid-oil slug flow over the water-oil case and up to 2.7 times higher heat transfer compared to SPF was observed in their investigation. Some recent studies have also examined the application of nanofluids in TBFs for heat transfer [50, 51]. Zhang et al. [50] conducted a numerical investigation to study CuO/water ($\phi \leq 3.0\%$ v/v)-nitrogen TBFs. A minor enhancement in the heat transfer compared to the water-nitrogen system was reported by them. Alrbee et al. [51] investigated the application of $Al_2O_3$/water nanofluid ($\phi \leq 1.0\%$ v/v) in TBFs and observed enhancement in heat transfer in the range of 10% - 65% compared to pure water case. A nominal increase (< 5%) in the pressure drop for $Al_2O_3$ nanofluid was also observed.

In our earlier investigation, it was demonstrated that the size of Taylor bubbles/unit-cells could be changed as well as a transition in the flow regime from the elongated Taylor bubbles to bubbly flow (bubble size smaller than channel diameter) could be achieved in a magnetically manipulated TBF of ferrofluids [39]. When the TBF of ferrofluids is exposed to a spatially distributed MF, a humped-shaped flocculate of ferrofluid gets formed due to the induced magnetic force, which creates a local magnetic pressure obstruction for the flow. The strategic manipulation of the magnetic pressure in the vicinity of the air-ferrofluid interface influences the mechanism of bubble formation, a detailed analysis of which is presented in the earlier article by the authors [39]. While our earlier work [39] showed the effect of magnetic manipulation on two-phase flow morphology, it was anticipated that such alteration would lead to the enhancement in two-phase heat transfer – the present work focusses on this hypothesis. A manipulation technique similar to our previous work [39] has been applied in the present study to change the size of generated Taylor bubbles/unit-cells (or flow morphology) at the given flow rates (or the Capillary number) of the respective phases, and their effect on thermal transport of the resulting two-phase flow is investigated. The present work provides an opportunity to examine the effect of elongated

Taylor bubbles as well as highly bubbly flow, and the effect of the liquid film thickness/void fraction on the heat transfer characteristics of the TBF of ferrofluids while geometrical and flow parameters are kept identical. Accordingly, the present study provides another unique perspective on the transport mechanisms, by taking the understanding derived from our previous work to the next level which includes heat transfer.

## 2. Experimental setup, materials and methodology

### 2.1 Experimental setup

A schematic and actual image of the experimental setup is presented in Figure 1. A T-junction mini-channel of a square cross-section of sides 3 mm ($D_h$ = 3 mm) is fabricated by the milling process on an acrylic substrate of dimensions: 500 mm (length) × 250 mm (width) × 12 mm (thickness). The T-junction forms the primary and orthogonal side flow channels. They are a popular way to generate segmented flows of two fluids [7, 34]. The inlet ports of the main and side flow channels are connected with the fluid-filled syringes. These syringes are driven by two different syringe pumps (New ERA® Model: 4000, Cole Palmer® Model: 78-9100C) for independent control on the flow rates of both phases. Ferrofluid (continuous phase) and air (dispersed phase) is pumped from the main and the side channel, respectively. Syringe pumps are used because of their precision in fluid dispensing, low flow rate requirement, and ease in handling ferrofluids. Pumps were calibrated before actual experiments through two independent methods [39] and found to be within the specified accuracy range.

For the heating section, a stainless steel (ss) strip (Dimensions: 150 mm (length) × 4 mm (width) × 0.07 mm (thickness)) is placed in such a way that the strip acts as the fourth wall of the channel as shown in the side view in Figure 1(ii). A DC power source (GW Instek, Model: SPS-606, Range: Voltage: 0-60 V, Current: 0-6 A) is used to supply constant power to heat the strip (Joule heating) during experiments to maintain a uniform heat flux boundary condition. Heat flux in the range of 6 - 9 kW/m$^2$ is supplied during experiments. The applied heat flux is kept constant in a given experiment. The heated section was insulated with glass wool to minimize environmental losses. The thickness of the heater strip is kept in micrometers to ensure that a negligible temperature gradient exists across the strip thickness (diffusion time scale of the heater is 1.46 × 10$^{-3}$ s), and the transient behavior of the flow is reflected instantly [16]. The heater section is placed at 110 mm downstream from the T-junction to provide sufficient length for the hydrodynamic development of the TBF. A slot has been made in the vicinity of the T-junction for the placement of the magnet close to the flow channel, as seen in the actual image (Figure 1(b)) of the setup.

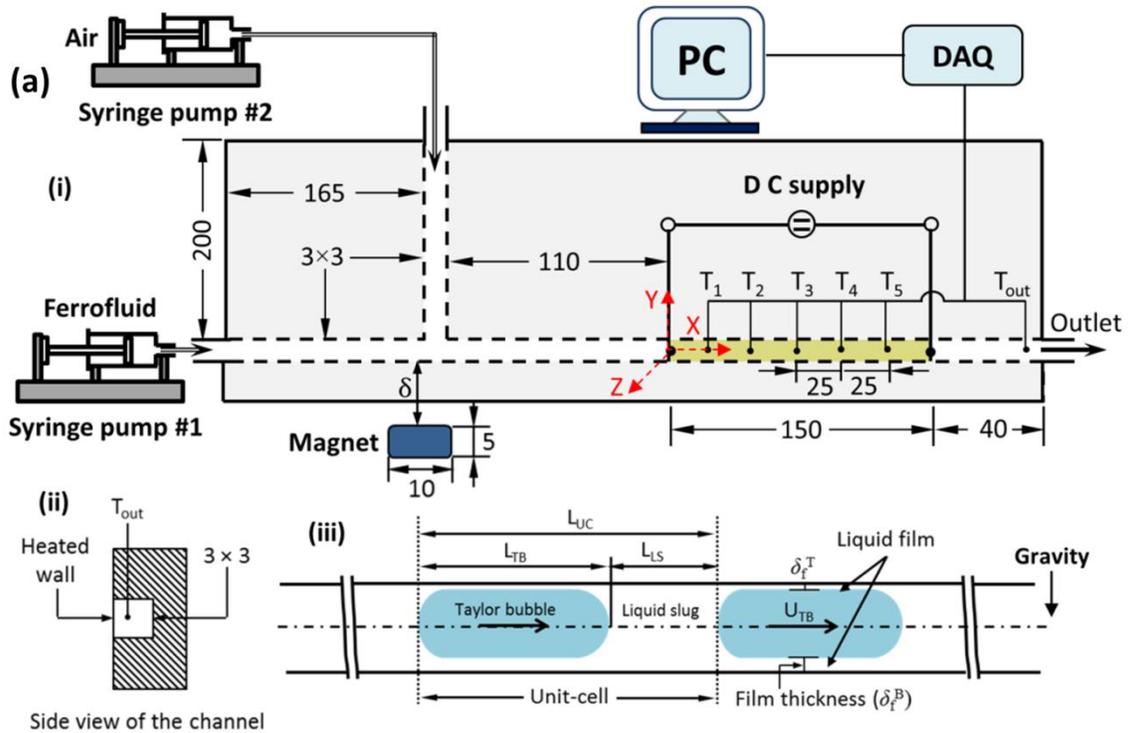

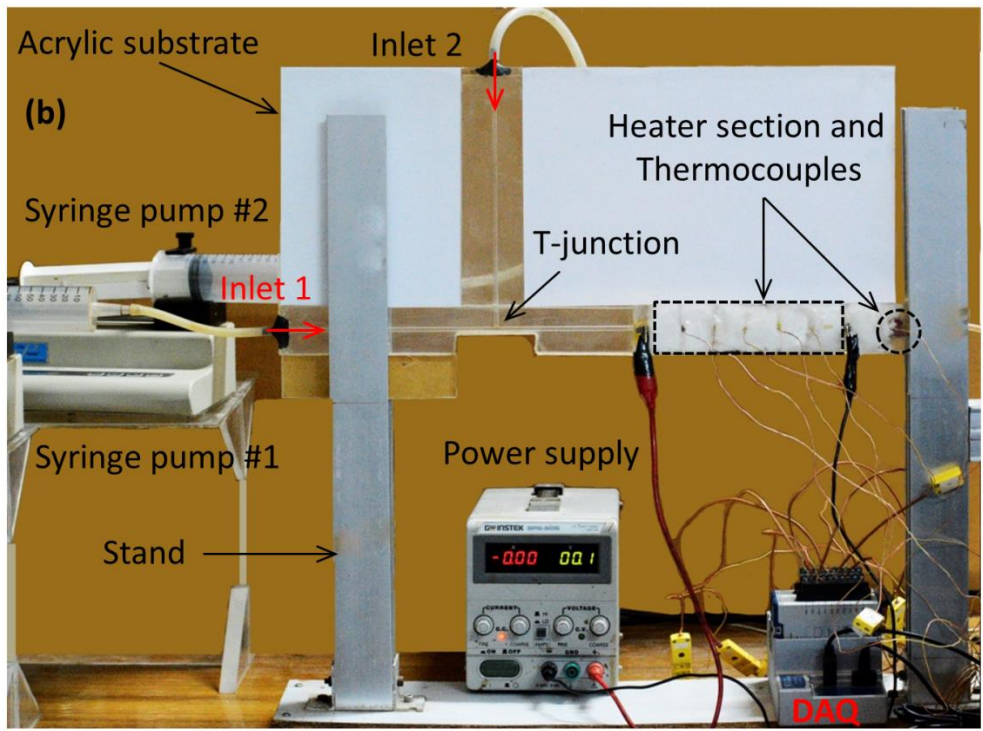

**Figure 1:** (a) (i) Schematic of the experimental setup with position of the magnet near T-junction (ii) Side view of the channel showing heated and insulated walls and the position of thermocouple near the outlet (iii) Flow morphology of TBF with relevant terminologies (b) actual image of the setup.

Five K-type thermocouples (bead diameter: 0.50 mm, Omega®) are placed on the heater strip ($T_1$-$T_5$, refer Fig. 1(i)) with the help of thermally conductive paste (OT 201-2, OMEGATHERM®) to measure the local wall temperatures at five different axial locations along the flow channel. Thermocouples are placed at a uniform distance of 25 mm from each other. The location of the first ($T_1$) and last ($T_5$) thermocouple is also at a 25 mm distance from the start and end of the heater, respectively. Another identical thermocouple is positioned adjacent to the flow outlet to measure the bulk fluid temperature at the outlet, as shown in Figure 1(ii). The inlet temperature of the fluid is measured at the start of each experiment. The data acquisition system (DAQ) of Nation Instruments® (cDAQ-9174, NI 9211) is used to acquire the temperature data during experiments. A DSLR camera (Model: Nikon 7100) with micro-lenses (Model: Nikkor 85 and 200 mm) is used for the image acquisition during experiments.

## 2.2 Materials

The aqueous ferrofluids of two different volumetric concentrations (0.25 and 0.50 % v/v) of IONPs are used in the present heat transfer experiments. The synthesis process and characterization of the ferrofluids are detailed in the earlier article [39]. Ferrofluids showed excellent stability throughout the investigations. Initial experiments were performed with water solution as the liquid phase for comparison and validation purposes. Due to the high surface tension of water at room temperature ($\sigma \sim 0.07$ N/m at 25°C), the generation of Taylor bubbles was observed to be aperiodic, forming different lengths of Taylor bubbles/unit-cells at the given flow rates. To generate the uniform size of Taylor bubbles periodically, the surface tension of water was lowered by mixing the Sodium dodecyl sulfate (SDS) surfactant at the critical micellar concentration (CMC, 2.4 gm/lit) [52]. The lowering of surface tension of water ensured uniformity (highly periodic) in the generated Taylor bubbles/unit-cells at all studied flow rates. In the case of ferrofluids, the generation of Taylor bubble and unit-cells were highly periodic with uniform size of bubbles and unit-cell due to their low surface tension. The thermo-physical properties of fluids used in the investigation are listed in Table 1. The thermo-physical properties of the dilute ferrofluids are estimated using standard correlations presented in the earlier articles [44, 46]. The magnetic properties of the ferrofluids are listed in Table 2. Properties are assumed to be constant as the rise in outlet fluid temperature is modest (5 to 15ºC from the inlet temperature) for causing any significant change in the fluid properties.

**Table 1: Thermo-physical properties (at 25ºC)**

| Fluid | Density, $\rho$ (kg/m$^3$) | Viscosity, $\mu$ (N-s/m$^2$) ×10$^{-4}$ | Surface tension, $\sigma$ (N/m) ×10$^{-3}$ | Specific heat capacity, $C_p$ (J/kg-K) | Thermal conductivity, k (W/m-K) |
|---|---|---|---|---|---|
| Air | 1.2 | 0.18 | -- | 1006 | 0.026 |
| Water | 998 | 8.9 | 34±2 | 4180 | 0.615 |
| FF #1 | 1008 | 9.2 | 21±1 | 4136 | 0.62 |
| FF #2 | 1017 | 9.5 | 22±1 | 4093 | 0.63 |

**Table 2: Magnetic properties**

| Ferrofluid | Nanoparticle volume fraction ($\phi$, %) | Saturation magnetization ($M_s$, [A/m]) |
|---|---|---|
| FF #1 | 0.25 | 697 ± 18 |
| FF #2 | 0.50 | 1461 ± 37 |

## 2.3 Methodology

The flow of gas and liquid phases meet at the T-junction, forming an interface, and a Taylor bubble-slug flow pattern is formed because of selected flow rates. Figure 1(iii) shows a schematic of the flow morphology of TBF with relevant nomenclature. As shown in the figure, bubbles are eccentrically located in the channel, leaving a higher thickness of liquid film near the bottom wall of the channel compared to the top wall ($\delta_f^B > \delta_f^T$) due to the effect of buoyancy. The effect of buoyancy force is observed for Bond number, Bo > 1 cases [53, 54]. In contrast, for Bo < 1, TBFs are dominated by the surface/interfacial tension force, and the liquid film is symmetrically distributed around the bubbles. For the present experiments, the Bond number lays in the range Bo ~ 3.5 – 4, showing the effect of buoyancy on bubbles in the resulting flow.

Liquids (water and ferrofluids) are used as the continuous phase, while air is used as the dispersed phase in the present experiments. The flow rate of water ($Q_{wa}$) is varied in the range = 2.5 - 15 ml/min ($Re_{wa}$ = 15 - 95), while the flow rate of air ($Q_{air}$) is kept constant at 10 ml/min ($Re_{air}$ = 3.67) in air-water experiments. As the flow rate of liquid is increased while flow of air is kept constant, both two-phase Reynolds number ($Re_{TP} = \rho_{ff} J_{total} D_h/\mu_{ff}$) and gas volume fraction ($\beta = Q_{air}/(Q_{air}+Q_{ff})$) vary simultaneously in the range (70 - 160) and (0.80 - 0.40), respectively. In the air - ferrofluid experiments, the formation of Taylor bubbles and unit-cell was uniform, having identical lengths with periodic formation at the given flow rates. In this case, too, the flow rate of air ($Q_{air}$) is kept constant at 10 ml/min ($Re_{air}$ = 3.55), and the flow rate of ferrofluids ($Q_{ff}$) is varied in the range of 2.5 - 10 ml/min ($Re_{ff}$ = 15 - 65). The $Re_{TP}$ and $\beta$ change in the range of 50 - 160 and 0.80 - 0.50, respectively. Based on the observations, additional experiments are also carried out in which flow rates of both air and ferrofluid is changed to keep $\beta$ in the range 0.75 - 0.80; the details of the same is provided in Section 4.2.1.

The local wall temperatures are acquired at five different axial locations during the experiments by the micro-thermocouples placed on the heater wall. The bulk fluid temperature at the outlet is also recorded by an identical micro-thermocouple during experiments, and the inlet temperature is measured at the start of each experiment. The variation in bulk fluid temperature along the axial locations is estimated by the linear interpolation of inlet and outlet fluid temperatures. While a non-intrusive direct measurement of bulk fluid temperature method is desired, it could not be achieved due to multiple challenges. Therefore, the linear interpolation approach is adopted from an earlier work of Mehta and Khandekar [16] to estimate the local bulk fluid temperature.

Magnet (Neodymium permanent magnet of cylindrical shape, Remnant magnetization, $B_r$ = 1.2 T) is placed in an upstream location (distance between the centers of the magnet and T-junction is 3.5 mm) in the vicinity of the interface forming at the T-junction. The placement location of the magnet is shown in Figure 1(a(i)). The placement of the magnet creates a spatial distribution of the MF, as shown in Section 4.1. Such MFs induce a magnetic force in ferrofluids. As demonstrated in the earlier article, the upstream/downstream placement of the magnet in the vicinity of the interface influences the mechanism of bubble formation [39]. When the magnet is placed upstream to the interface, it causes early shearing of the bubbles resulting in the generation of smaller sized bubbles. For a given strength of the MF, it was also observed that the upstream placement of the magnet generated the smallest sized bubbles compared to all other tested locations of the magnet placement in the vicinity of the interface [39]. Therefore, in the present experiments, the magnet is placed in the same upstream location to produce its maximum effect on the resulting flow morphology. The distance of the magnet from the flow channel is varied (in the range 3 mm - 10 mm from the closest wall of the channel) to change the strength of the applied MF while its upstream location with respect to the interface is kept unchanged.

The induced magnetic force in ferrofluid is changed either by varying the strength of the applied MF, as explained above or by varying the volumetric concentration of IONPs nanoparticles (FF #1: 0.25% v/v and FF #2: 0.50% v/v). The wall and outlet bulk fluid temperatures are acquired at 20 and 30 Hz for air-water and air-ferrofluid TBF experiments, respectively, after a quasi-steady state is reached (TBFs are inherently transient in nature). The acquired quasi-static temperature data is time-averaged over 900 - 1500 data points for the heat transfer analysis.

## Numerical simulation of the magnetic field (*H*)

The measurement of the spatial distribution of the MF and its gradient created by the magnet is relatively difficult to carry out experimentally. Therefore, numerical simulations are performed to quantify and visualize the spatial distribution of the magnetic field (*H*), its gradient (∇*H*) and magnetic flux density (*B*) generated by the magnet in the domain of interest. Maxwell equations are solved on the COMSOL Multiphysics® (V 5.3a) platform to compute the distribution of *H*, ∇*H* and *B* in the domain of interest. Following partial differential equations are solved to simulate the MF

**Gauss law:** $\nabla \cdot B = 0$, **Ampere's Law**: $\nabla \times H = 0$ (1)

**Magnetic induction:** $B = \mu_0 \mu_r H + B_r$ (2)

Flux insulation boundary condition ($\hat{n} \cdot B = 0$, where $\hat{n}$ is the unit vector normal to the surface) is applied to the domain boundaries for the simulation. Remnant magnetization ($B_r$) and relative permeability ($\mu_r$) of the Neodymium magnet is defined for the simulation of the MF. Three-dimensional simulation is performed for the MF, and the domain of interest is discretized with third-order (cubic) tetrahedral mesh elements. Validation of the simulation results for the *B* field created by the magnetic is performed with the analytical solution, and the same is presented in Section 4.1. A similar numerical simulation was performed to obtain magnetic field and magnetic force details in our previous work [39]. More details on the simulation domain, procedures of the numerical simulation and validation of the MF results can be found in the previous articles of the authors [39, 46].

## Data reduction

The quasi-static wall/fluid temperature is time-averaged and normalized as follows

$$\bar{T}_{wall/fluid} = \frac{\sum_1^N T_{wall/fluid}}{N} ; T^*_{wall/fluid} = \frac{\bar{T}_{wall/fluid} - \bar{T}_{fi}}{(q^{"}D_h)/k_{ff}} \tag{3}$$

where, N is the number of sample points, $q^{"}$ (applied heat flux) and $T_{fi}$ (inlet fluid temperature) are known parameters in any given experiment.

The time-averaged local heat transfer coefficient ($h_{local}$) and local Nusselt number ($Nu_{local}$) along the flow direction are defined as:

$$h_{local} = \frac{q^{"}}{(\bar{T}_w^n - \bar{T}_f^n)} ; Nu_{local} = \frac{q^{"}D_h}{(\bar{T}_w^n - \bar{T}_f^n)k_{ff}} \tag{4}$$

where, $\bar{T}_w^n$ = Time-averaged local wall temperature acquired by the thermocouples placed on the heater at different axial locations (indicated by $n$ = 1, 2..5), $\bar{T}_f^n$ = local fluid temperature along the axial direction, which is obtained by linear interpolation of the inlet and outlet fluid temperatures. The $h_{local}$ and $Nu_{local}$ is calculated from the time-averaged local temperature data at five measured locations. The average heat transfer coefficient ($h_{avg}$) and average Nusselt number ($Nu_{avg}$) is computed by spatial averaging of local values. Uncertainty analysis has been carried out to ascertain the possible errors in the measurements and computations of the present work, as presented in the appendix section. The maximum uncertainties in the computation of $T^*_{wall}$, $h_{local}$, and $Nu_{avg}$ is in the range of ±10%. Thermocouples were calibrated before experiments using a temperature bath (Thermo-Haake® DC10-K20; ±0.1ºC, with RTD PT-1000 NIST traceable calibration) and found to be within the specified uncertainty range (±0.3ºC). Sample calibration curves are provided in the appendix.

## 3. Validation of the experimental results

Validations of the present work are carried out with the experimental results of previous studies as well as with the prediction of empirical correlations proposed by different authors for the estimation of heat transfer in TBFs. The local and average Nusselt number ($Nu_{local}$, $Nu_{avg}$) calculated for the present study is compared with the relevant previous studies. In the first part, $Nu_{local}$ computed for the present work is compared with the experimental results of Majumdar et al. [14] because of similar experimental setup, material properties and boundary conditions. An experimental study was conducted by them to investigate the heat transfer characteristics of air-water TBF in a square cross-section mini-channel of sides 3.3 mm. A uniform heat flux boundary condition was applied on the bottom wall of the channel while the other three walls were kept insulated, similar to the present case. In their experiments, both bulk fluid and wall temperatures were measured at three different axial locations using thermocouples (intrusive micro-thermometry for the bulk fluid temperature). The $Nu_{local}$ was then computed from the measured wall and fluid temperatures and applied heat flux boundary conditions at the three axial locations for different gas volume fractions ($\beta$) and plotted against the inverse Graetz number ($x^* = x/Re \cdot Pr \cdot D_h$). For the present study, the bulk fluid temperature at different axial locations is estimated by the linear interpolation of the inlet and outlet fluid temperatures as stated earlier to avoid intrusive measurement, which could alter/modify the flow locally. Figure 2(a) presents the comparison of $Nu_{local}$ plots against $x^*$ obtained for the present experiments with the results of Majumdar. The $Nu_{local}$ plot is also compared with the experimental observations of Walsh et al. [18] (They had performed heat transfer experiments on TBFs in a cylindrical capillary tube ($D_h$ = 1.5 mm) under uniform heat flux boundary condition applied to the circumference). As seen from the figure, our experimental results lay well within the observations (± 20%) of the earlier studies.

The second part of validation is carried out with the predictions of two different correlations proposed by Leung et al. [17] and Walsh et al. [18] for the estimation of average Nusselt number ($Nu_{avg}$) in TBFs. The correlations of Leung [17] and Walsh [18] are based on the extension of Gratz theory for the thermally developing laminar convective single-phase flows in tubes. As liquid phase is regarded as the major contributor in the heat exchange process in TBFs, Gratz theory is generally used to model the heat transfer in such flows. The $Nu_{avg}$ computed for the present experiments are compared with the predictions of the correlations for identical flow conditions and morphology. Figure 2(b) shows the comparison of plots of the $Nu_{avg}$ against two-phase Reynolds number ($Re_{TP}$) and the corresponding gas fractions ($\beta$) for the air-water TBF in the bottom wall heating case. The $Nu_{avg}$ computed for the present experiments lies within the predictions of the two correlations, as seen from the figure. Here, it is to be noted that the observed deviations can be attributed to the fact that the correlations are developed based on the Gratz solution for thermally developing laminar single-phase flow in cylindrical tubes with circumferential heating, for which the Nusselt number in fully developed regions is 4.36 [55]. Leung [17] and Walsh [18] developed their correlations with similar experiments conducted in cylindrical tubes with circumferential heating. However, in the present case, only one wall is heated while the other three walls are insulated. Therefore, the $Nu_{avg}$ in the present case is expected to differ from the predictions. The discrepancies in the prediction of Nusselt number for two-phase TBF/slug flow by different correlations (for identical conditions) have also been pointed out by other authors as well [9, 56]. Such discrepancies are possibly emanating because heat transfer in slug flows is highly dependent on the flow morphology or phase distribution, and those parameters can vary locally in different experiments. The validation of the present work builds necessary confidence in the experiments.

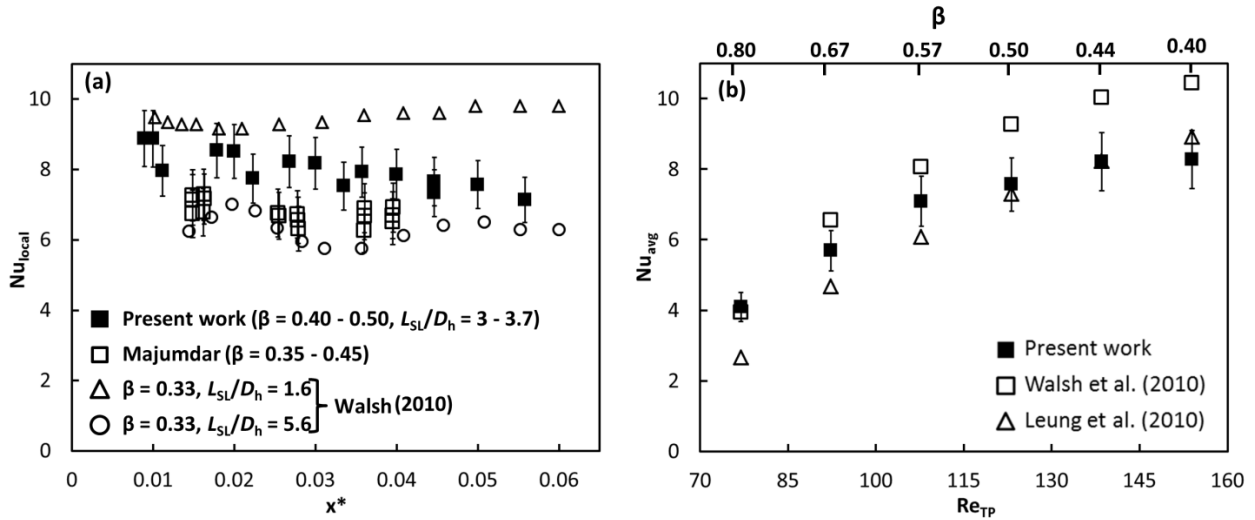

**Figure 2:** (a) Comparison of local Nusselt number ($Nu_{local}$) vs inverse Graetz number ($x^*$) computed for the present experiments with the result of Majumdar et al. [14] and Walsh et al. [18] (b) Comparison of the average Nusselt number ($Nu_{avg}$) computed for the present experiments with the prediction of Leung et al. [17] and Walsh et al. [18] correlations for the air-water TBF.

## 4. Results and discussion

The flow of two immiscible phases (air and ferrofluid) meets at the T-junction, and a Taylor bubble train flow pattern is formed in the resulting two-phase flow. The flow morphology of the TBF is modified by the application of a magnetic field. The mechanism of bubble formation in T-junction geometries can be influenced through external magnetic manipulations. Such manipulations provide control over the size of generated Taylor bubbles/unit-cells in the resulting two-phase flow [39]. Smaller sizes of Taylor bubbles/unit-cells are generated by varying the intensity of the induced magnetic force while flow rates are kept unchanged. To ensure mass conservation, the frequency of bubble generation increases as the size of generated bubbles/unit-cells gets smaller.

Figure 3 shows the effect of magnetic manipulation on the TBF of air - FF #2 ferrofluid when the magnet is placed at the upstream location in the vicinity of the T-junction, as detailed in the methodology section. Following observations can be made from the acquired images as seen from the figure, (i) Bubbles are eccentrically located in the flow because of the buoyancy effect, which leaves a higher thickness of liquid film on the bottom wall compared to the top wall for all cases (ii) Both size and shape of the generated bubbles changes due to magnetic manipulation and smaller sizes of bubbles and unit-cells are produced in MF compared to no-field cases (iii) Different sizes of bubbles (or phase distribution) can be generated at same flow rates by varying the intensity of the applied MF (iv) The void fraction of the resulting flow decreases and film thickness near the bottom wall ($\delta_f^B$) increases with the generation of smaller sized bubbles (v) Liquid film thickness at the top wall ($\delta_f^T$) is nearly identical for all cases due to buoyancy effect. The changes in the length, $L_{TB}^* = (L_{TB}/D_h)$, $L_{UC}^* = (L_{UC}/D_h)$ and film thickness ($\delta_f^* = (\delta_f^T$ or $\delta_f^B)/D_h$) with the MF are plotted in Figure 3(c). The film thickness at the bottom wall ($\delta_f^B$) increased by 40% at $B = 0.19$ T from the no-field case, as seen from the plots in Figure 3(b). The film thickness near the top wall is constant ($\delta_f^* \sim 0.13$) for all the cases. The effect of gravity for large Bond number (Bo > 1) cases have also been observed and investigated in earlier studies [53, 54]. Maximum uncertainties in the measurement of $L_{TB}^*$, $L_{UC}^*$, $\delta_f^*$, and $f$ is in the range of ± 8%.

It is evident from observations that the application of MF significantly alters the flow morphology of the TBFs of ferrofluids. For pure fluids, such changes in the size of bubbles/unit-cells and the film thickness are usually achieved by increasing the Capillary number (Ca) of the flow, which changes the mechanism of bubble formation from squeezing (Ca ~ $10^{-4}$ - $10^{-3}$) to transitional (Ca ~ $10^{-3}$ - $10^{-2}$) and dripping (Ca ≥ $10^{-2}$) regime in T-junction geometries [34]. It can also be altered through other means, as discussed in the introduction. However, such an alteration is achieved by the application of an MF in the present case without changing the Ca of the flow. Smaller bubbles/droplets are formed in transitional (balance of interfacial and shearing forces) and dripping flow regimes compared to squeezing (interfacial tension dominated) regimes [34]. The thickness of the liquid film increases (as $Ca^{2/3}$), and the intensity of flow circulation gets weaker with an increase in the Ca [2, 19, 57].

As stated earlier, the transport characteristics and ensuing coefficients of the TBFs/Slug flows are primarily dependent on the phase distribution (length of slugs/unit-cells) and film thickness/void fraction. As these parameters get affected due to magnetic field application, it is expected that the heat transfer characteristics would be different for cases in which heater works as the top, bottom, and side wall of the channel, as the liquid film distribution is asymmetric around the bubbles for NMF and MF cases. To examine all such cases, the orientation of the heater wall was changed to make it work as the bottom, top, and side wall of the channel in different experiments. The effect of the flow morphology and liquid film on the heat transfer characteristics of TBFs have been analyzed and discussed for different cases in the present work.

## 4.1  Magnetic field ($H$), its gradient ($\nabla H$) and magnetic flux density ($B$)

Before proceeding further, results for the MF simulation are shown in the present section. Simulations provide useful insights into the spatial distribution of the $H$ field and its gradient ($\nabla H$) created by the magnet, which is difficult to gain experimentally. The numerical simulation also provides important information about the effective spatial range of the MF and induced magnetic force produced by the magnet. The simulation results for the $H$ field, its gradient ($\nabla H$), and magnetic flux density ($B$) are shown in Figure 4(a, b). Spatial distribution of the $H$ and $\nabla H$ is plotted (Figure 4(a)) in the domain of interest (fluidic channel) up to 25 mm on each side from the center of the magnet when it is placed 3 mm below the bottom wall of the channel. The plot is on the central sectional place ($x$-$y$ plane at $z = 0$, refer inset of Figure 4(b) for the coordinate system). It is clear from the plots that the field and its gradient are high in the close range (up to 10 mm). The maximum strength of the applied MF is $1.5 \times 10^5$ A/m, which translates to the magnetic flux density ($B = \mu_0 H$) of 0.19 T and maximum of the gradient ($\nabla H$) is $4.6 \times 10^7$ A/m$^2$, which produces a maximum induced force of ~ $3.2\times10^{10}$ N/m$^3$ in FF #1 and ~ $6.7\times10^{10}$ N/m$^3$ in FF #2 ferrofluids. Flux density ($B$) generated by the magnet along the cylindrical axis ($z$) is plotted up to 15 mm distance from the magnet and matched with the analytical solution for the same (Eq. 5). Excellent agreement between simulation results and analytical solution can be seen in Figure 4(b), which also validates the numerical approach. Analytical solution for the $B$ field created by a finite size cylindrical permanent magnet along its cylindrical axis is given by [58]

$$B = \frac{B_r}{2}\left(\frac{T+y}{\sqrt{R^2+((T+y)^2)}} - \frac{y}{\sqrt{R^2+y^2}}\right) \quad (5)$$

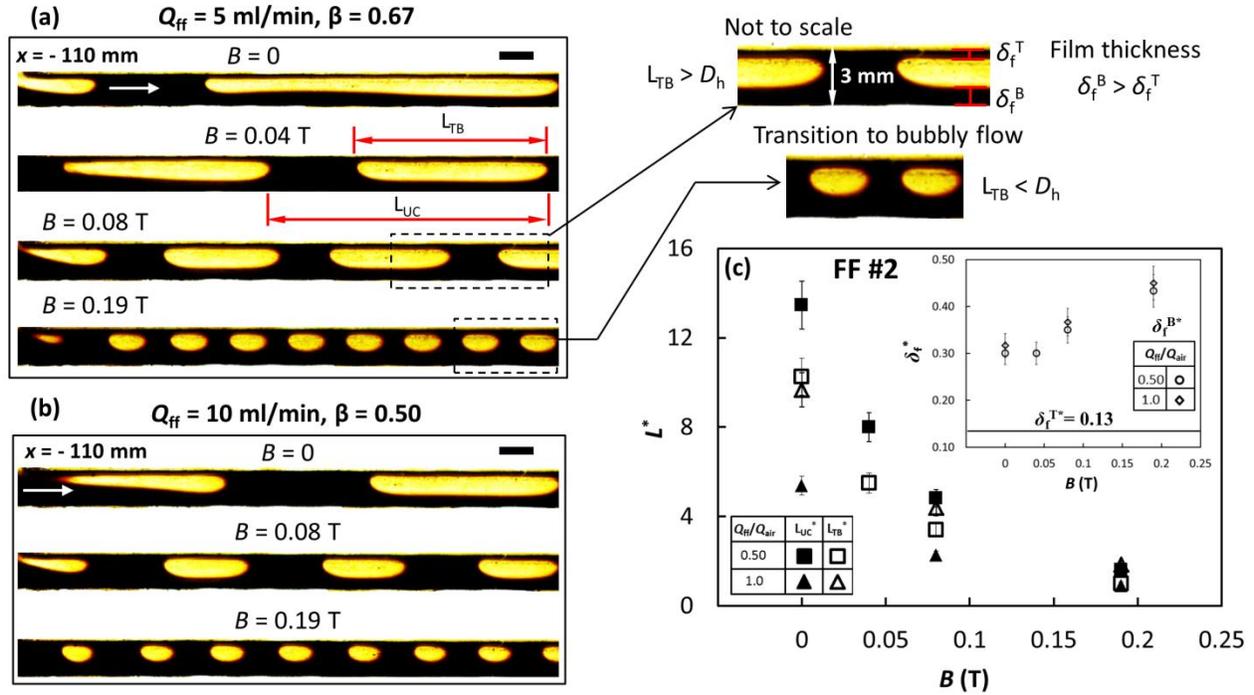

**Figure 3:** Effect of MF on the TBF at (a) $Q_{ff}/Q_{air}$ = 0.50, $\beta$ = 0.67 (b) $Q_{ff}/Q_{air}$ = 1.0, $\beta$ = 0.50; Applied flux density ($B$) is changed from 0 to 0.19 T by changing the distance ($\xi$) of the magnet from the lower wall of the channel (for FF #2, length of scale bars: 3 mm ) (c) Comparison plots of normalized length of Taylor bubbles ($L_{TB}^*$) and unit-cells ($L_{UC}^*$) and (inset) Film thickness ($\delta_f^*$) near the bottom and top walls against the applied field strength ($B$).

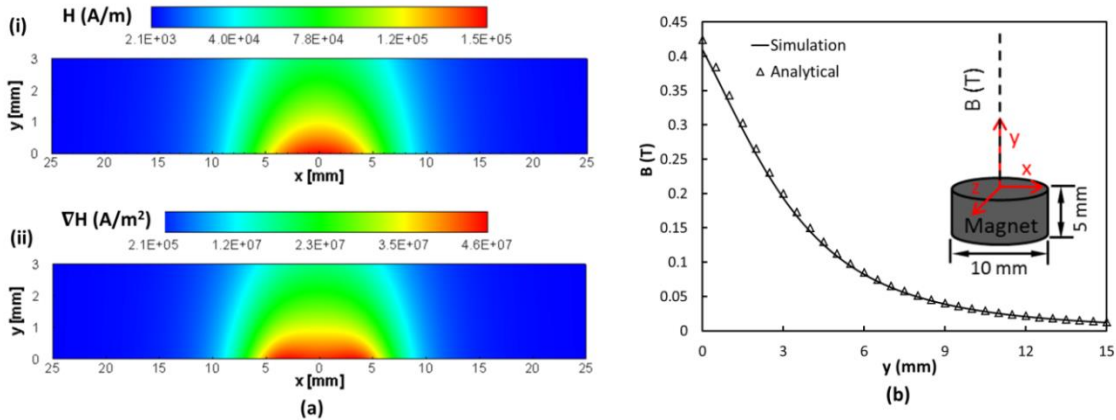

**Figure 4:** (a) Spatial distribution of (i) $H$ field and (ii) its gradient ($\nabla H$) produced by the magnet (plots are on the *x-y* central sectional plane at *z* = 0, refer to coordinate system shown in the adjacent figure) plotted up to 25 mm on either side of the center of the magnet in the domain of interest (center of the magnet is at *x* = 0, magnet is placed at 3 mm distance from the bottom wall of the sectional plane in symmetrical orientation with respect to the plane) (b) $B$ field induced by the magnet, simulation results are compared with the analytical solution.

Magnet is placed at three different locations (3 mm, 6 mm, and 9 mm) from the closest wall of the flow channel, which created a maximum flux density of 0.19 T, 0.08 T, and 0.04 T, respectively, in magnetic field experiments.

### 4.2 Thermal transport in magnetically manipulated TBF

The formation of Taylor bubbles are due to the build-up of squeezing pressure in liquid flow upstream to the interface in the absence of the MF for investigated range of Capillary number, Ca$_{\text{ff}}$ = (1.9 − 8.0) × 10$^{-4}$. When a magnet is placed at the upstream location in the vicinity of the interface, it affects the mechanism of bubble formation. The effect of the MF on TBF of air and FF #1 is shown in Figure 5 (FF #2 case is shown in Figure 3) at different flow rate ratios ($Q_{\text{ff}}/Q_{\text{air}}$) and the corresponding gas fractions ($\beta$). The upstream location of the magnet with respect to the interface and its distance from the lower wall of the channel is also shown in Figure 5(a). The influence of induced magnetic force on the mechanism of bubble formation and a detailed analysis of the resulting flow morphology is presented in our earlier article [39]. As seen in Figures 3, 5 and 6, the size of the generated bubbles and unit-cells gets smaller with a corresponding increase in their generation frequency as the intensity of the applied field is increased from 0 to 0.19 T. It is also seen that the distribution of the liquid film is asymmetric having a higher amount of liquid flow near the bottom wall of the channel than the top wall. As the flow rate of ferrofluids is increased (from 2.5 ml/min to 10 ml/min), the size of bubbles gets smaller in the NMF cases, too, as seen in the figures. A validation of bubble and unit-cell length observed in the present NMF cases with results of Thulasidas et al. [59] is shown in the inset of Figure 5(d). Thulasidas et al. [59] fitted their experimental observations for bubble and unit-cell lengths in the form of a correlation as $L_{\text{TB}}/L_{\text{UC}}$ = 1 - ($Q_{\text{ff}}/Q_{\text{total}}$) relating ratio of bubble and unit-cell length with flow rates. The ratio of bubble and unit-cell length of present cases is in close agreement with the correlation, as seen in the figure.

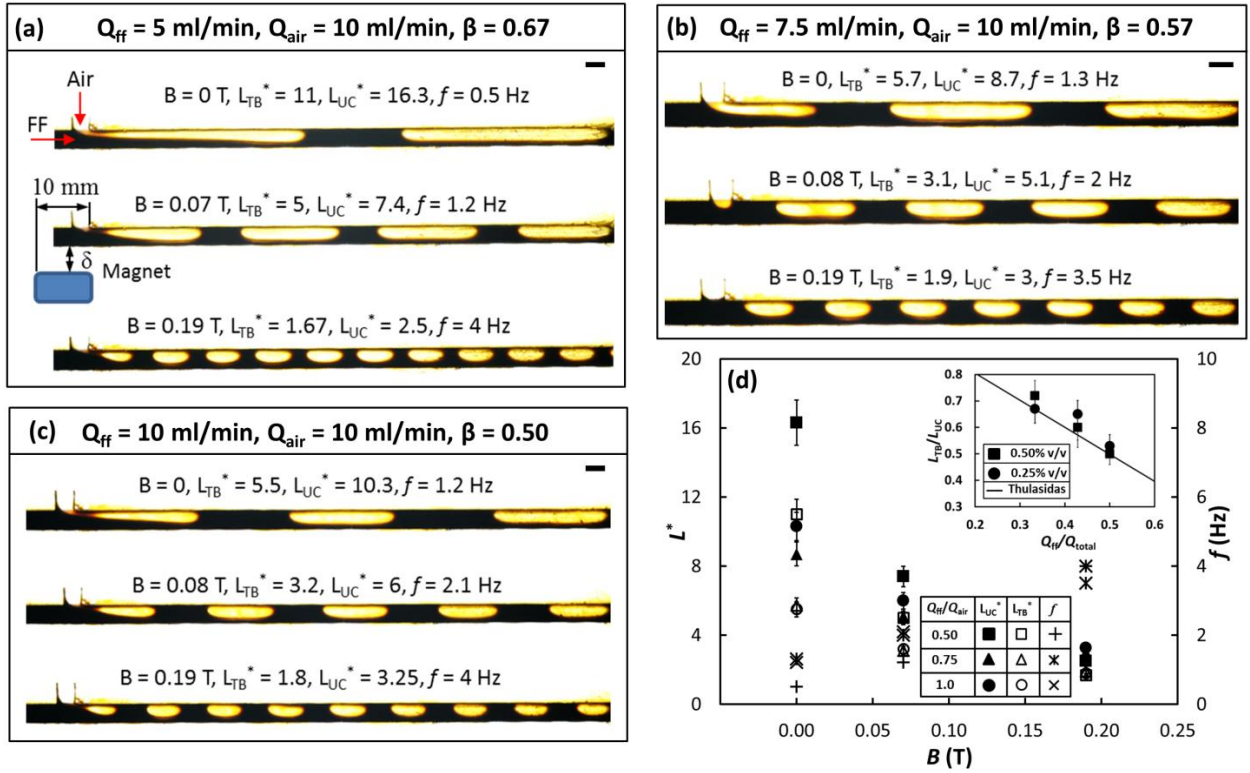

**Figure 5:** Effect of the MF on TBF of air and FF #1 at different volume flow rates and gas fractions ($\beta$), Images of resulting TBF morphology and normalized length of Taylor bubbles ($L_{TB}^*$) and unit-cells ($L_{UC}^*$) and generation frequency of Taylor bubbles ($f$) at flow rate ratio ($Q_{ff}/Q_{air}$) of (a) 0.50 (b) 0.75 (c) 1.0 (length of scale bars: 3 mm) and (d) plots of $L_{TB}^*$, $L_{UC}^*$ and $f$ against the flux density ($B$) and (inset) A validation of observed bubble and unit-cell length in NMF cases: Plots of $L_{TB}/L_{UC}$ vs $Q_{ff}/Q_{total}$ for both ferrofluids compared with the correlation of Thulasidas et al. (1995).

A comparison of the size of Taylor bubbles and unit-cells and their generation frequency at $Q_{ff}$ = 2.5 ml/min, $Q_{air}$ = 10 ml/min, $\beta$ = 0.80 for FF #1 and FF #2 ferrofluids is shown in Figure 6 at $B$ = 0, 0.05 T and 0.19 T. Due to higher concentration of IONPs in FF #2 (0.50% v/v), the induced magnetic force will approximately be double of FF #1 (0.25% v/v). This is because the magnetic force is proportional to the concentration of IONPs in ferrofluids [39]. A higher magnetic force also produces a greater effect on the mechanism of bubble formation and causes early shearing and pinch-off of the bubbles [39]. This result in the generation of smaller sized bubbles and unit-cells, which can also be seen in the present case. As seen from Figure 6, the size of bubbles/unit-cells are smaller for FF #2 compared to FF #1 at the given field strengths. A transition in the flow regime from elongated Taylor bubbles to bubbly flow (bubble length smaller than channel diameter) can also be seen for FF #2 at 0.19 T (Figure 6(b(iii))). Bubble lengths bigger than three times that of the channel diameter ($L_{TB}^* > 3$) are referred to as 'larger bubbles', and those smaller than that value are referred to as 'smaller bubbles' in this article.

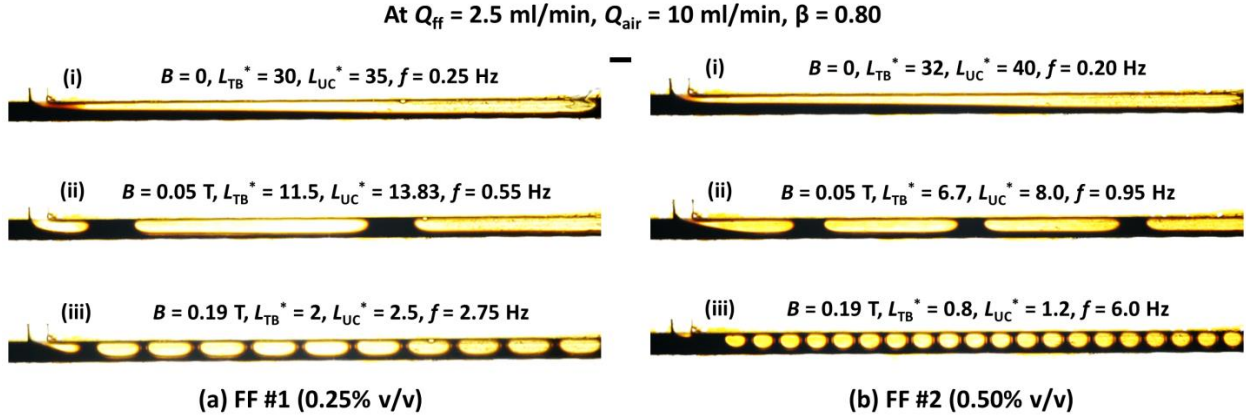

**Figure 6:** Comparison of $L_{TB}^*$, $L_{UC}^*$, and $f_{TB}$ and resulting TBF morphology for (a) FF #1 and (b) FF #2 at $Q_{air}$ = 10 ml/min and $Q_{ff}$ = 2.5 ml/min at $B$ of (i) 0, (ii) 0.05 T and (iii) 0.19 T (scale bar represents 3 mm length).

A comparative analysis of the heat transfer characteristics of the magnetically manipulated TBFs of FF #1 and FF #2 in the case of heater working as the bottom wall of the channel is presented in Figures 7 and 8 at the homogeneous gas fractions ($\beta$) of 0.80 and 0.67, respectively. A time-averaged analysis of the acquired quasi-static temperature data is carried out to compute the local and spatially averaged two-phase heat transfer coefficient ($h_{local}$, $h_{avg}$) for both NMF and MF cases. The time-averaged ($\bar{T}_{wall}$), normalized wall temperatures ($T_{wall}^*$) and $h_{local}$ are plotted against the normalized axial length ($x/D_h$) of the heater. The $h_{avg}$ and averaged Nusselt number ($Nu_{avg}$) are plotted against the applied flux density ($B$) and compared for both the ferrofluids. The plots are at the same Re and Ca but at different $B$ fields. The linear interpolation of the fluid temperature ($\bar{T}_{fluid}$, $T_{fluid}^*$) is also shown in Figures 7(b(i)). The plots of $\bar{T}_{wall}$ and $T_{wall}^*$ in Figures 7(a(i), b(i)) and $\bar{T}_{wall}$ in Figures 8(a(i), b(i)) show that the wall temperatures decrease when an MF is applied as the flow morphology changes from longer bubbles in NMF cases to smaller sized bubbles/unit-cells in the MF cases. For example, the $\bar{T}_{wall}$ or $T_{wall}^*$ is ~ 25% and ~ 20% lower at $B$ = 0.19 T compared to NMF ($B = 0$) for the FF #2 ferrofluid at flow rate of 2.5 and 5 ml/min, respectively, as seen in Figures 7(b(i)) and 8(b(i)). Similarly, fluid temperature ($T_{fluid}^*$) at exit also increases, as seen in Figure 7(b(i)).

The observed changes in the wall and fluid temperatures indicate an improvement in the heat exchange process for smaller bubble/unit-cell cases. The plots of $h_{local}$, $h_{avg}$ and $Nu_{avg}$ in Figures 7 and 8 reveal these enhancements in the heat transfer. An augmentation of more than 100% in the heat transfer is seen in Figures 7(b(ii, iii)) at $B = 0.19$ T compared to the NMF case when flow morphology changes from the longer bubbles having a length scale equivalent to the heater to spherical bubbles having a length smaller than channel diameter (refer Figure 6). The enhancement in the heat transfer is higher for FF #2 (~ 107%) compared to FF #1 (~ 70%) at $B = 0.19$ T and $\beta = 0.80$ because of comparatively smaller bubbles/unit-cells are generated in air - FF #2 case, as seen in Figure 6. As the flow rate of ferrofluids is increased to 5 ml/min ($\beta = 0.67$), the size of generated bubbles reduces even for the NMF case compared to the previous case of 2.5 ml/min. A higher flow of liquid increases heat transfer for both NMF and MF cases; however, enhancement due to the application of an MF reduces as the size of generated bubbles gets smaller for NMF cases. The net enhancement reduces to ~ 40% for FF #1 and ~ 55% for FF #2 at 0.19 T from the corresponding NMF cases. It can also be seen from the plots that by varying the intensity of the induced magnetic force, different sizes of Taylor bubble/unit-cells can be formed, and the net enhancement in heat transfer can be adjusted between an upper and lower limit through such manipulations, as seen in Figures (7, 8) (b(iii), c(iii)).

The models for heat transfer in TBFs/slug flow have been proposed by multiple authors, and the liquid phase has been recognized as the primary contributor to the overall heat exchange process in all such models [4, 17, 18, 30, 32]. As per various authors, heat is first transferred to the deposited liquid film, which is stagnant and works as a conductive medium. Through conduction from the film, heat is taken away by the periodic motion of the liquid slugs and gas bubbles with circulating flows. The contribution of the gas phase is minimal in the overall heat transfer process as the thermal properties of the gas phase are inferior to the liquids. An illustration of the heat transfer process in TBF is presented in Figure 9(a), with red arrows indicating the flow of heat. The figure also shows the formation of close streamlines in a liquid slug and its effect on temperature distribution, as reported by Bajpai and Khandekar [60]. It can be seen that the flow circulation bring colder fluid from the bulk to heated zones which improves heat exchange. It has also been observed that the smaller bubbles/unit-cells are better for higher heat transfer in previous studies, as discussed in the introduction. Leung et al. [54] investigated the effect of gravity in two-phase heat transfer in horizontal TBFs in a circular tube ($D_h$ = 2.12 mm). Drainage of flow from the top to the bottom side in gravity dominated flows (Bo ~ 1.03) was observed by them. The thickness of liquid film near the bottom wall was slightly higher compared to the top wall. However, little effect of buoyancy on two-phase heat transfer was observed in their cases. The heat transfer was similar in vertical and horizontal orientations of the heated tube.

The observed enhancement in the heat transfer in MF cases can be attributed to the two major modifications caused by the MF in the resulting two-phase flow. The gas void fraction of the resulting TBF is smaller at any given cross-sectional area because of the formation of smaller sized bubbles compared to larger bubbles in NMF cases. For example, the liquid film thickness increased by ~ 40% in MF cases compared to NMF cases, as seen in Figure 3. This means the cross-sectional void fraction has reduced by ~ 40% in such flows, and the interaction of the liquid phase with the heated surface has increased by a proportional amount (as the film thickness is constant near the top wall in all cases). Such modifications result in a higher amount of liquid participating in the heat exchange process in MF cases compared to the NMF cases. It can also be deduced from the visual observations that the liquid film near the bottom wall is not stagnant as in surface tension dominated cases because a higher thickness of the stagnant film would have presented an additional diffusion resistance to the heat flow, which should have resulted in a decrease in the heat transfer for MF cases. The thick film indicates convection currents due to fluid motion between bubbles and the bottom wall, which contributes to the overall enhancement in heat transfer. It can also be observed that a higher number of smaller unit-cells are present in the heated zone at any time instance, in MF cases over NMF. This creates multiple fluid mixing sites caused by the circulating flows, further improving the heat exchange process. Figure 9(b) illustrates the expected heat transfer mechanisms in the magnetically manipulated TBFs. The effect of magnetic manipulation in other heater wall orientation cases is analyzed and compared in the next section.

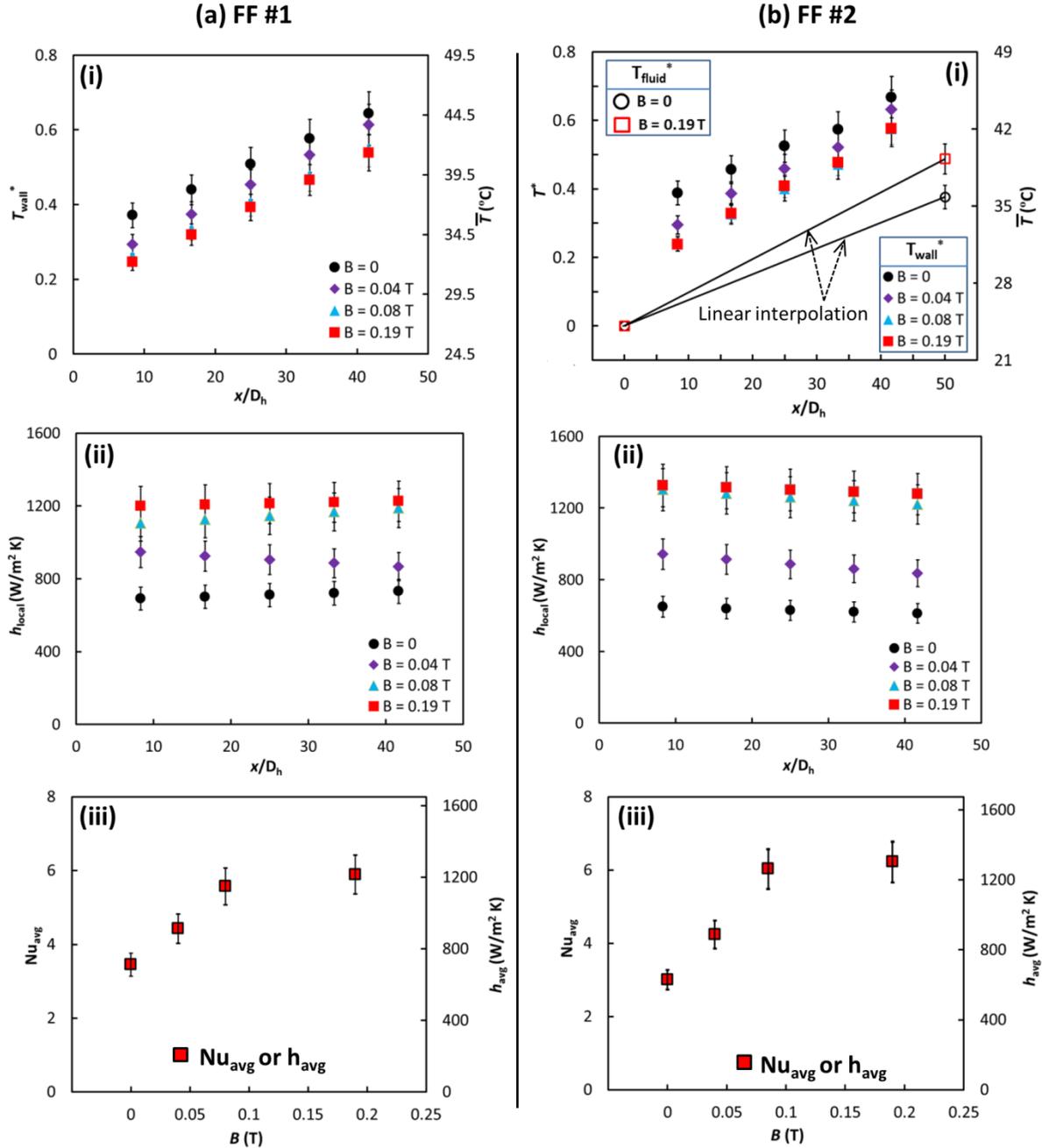

$Q_{ff}$ = 2.5 ml/min, $Q_{air}$ = 10 ml/min, β = 0.80, Heater as Bottom wall

**Figure 7:** Comparison of (i) normalized ($T^*_{wall}$) and time-averaged ($\overline{T}_{wall}$) wall temperature with linear interpolation of fluid temperature and (ii) local convective heat transfer coefficient ($h_{local}$) against the normalized axial length ($x/D_h$); (iii) Averaged Nusselt number ($Nu_{avg}$) and convective heat transfer coefficient ($h_{avg}$) at different magnetic strength (*B*) for (a) FF #1 (b) FF #2 at β = 0.80.

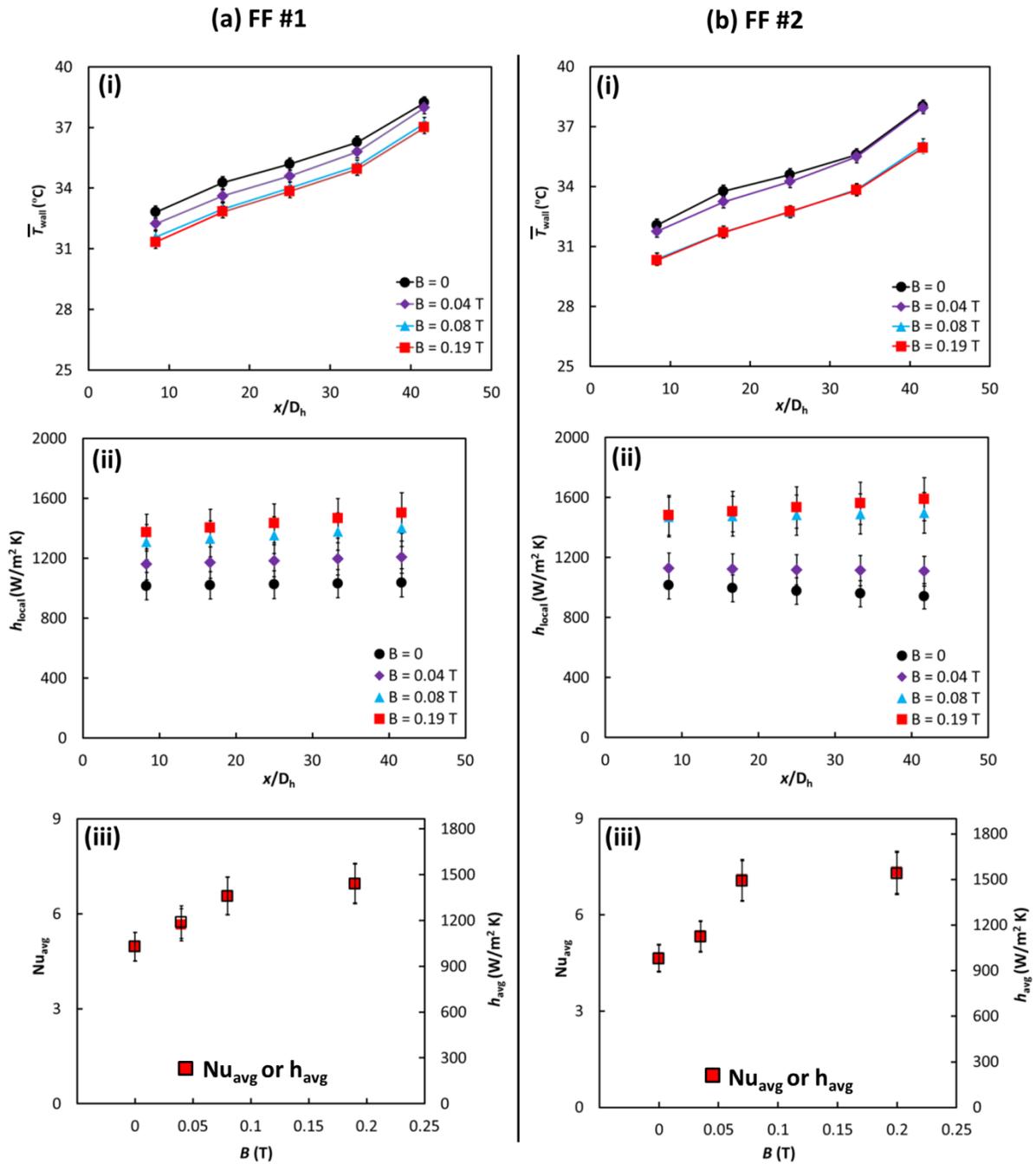

$Q_{ff}$ = 5 ml/min, $Q_{air}$ = 10 ml/min, β = 0.67, Heater as bottom wall

**Figure 8:** Comparison of plots of (a) FF #1 (b) FF #2 for (i) $\overline{T}_{wall}$, (ii) $h_{local}$ against the ($x/D_h$) and (iii) $Nu_{avg}$ and $h_{avg}$ at different magnetic strengths at β = 0.67.

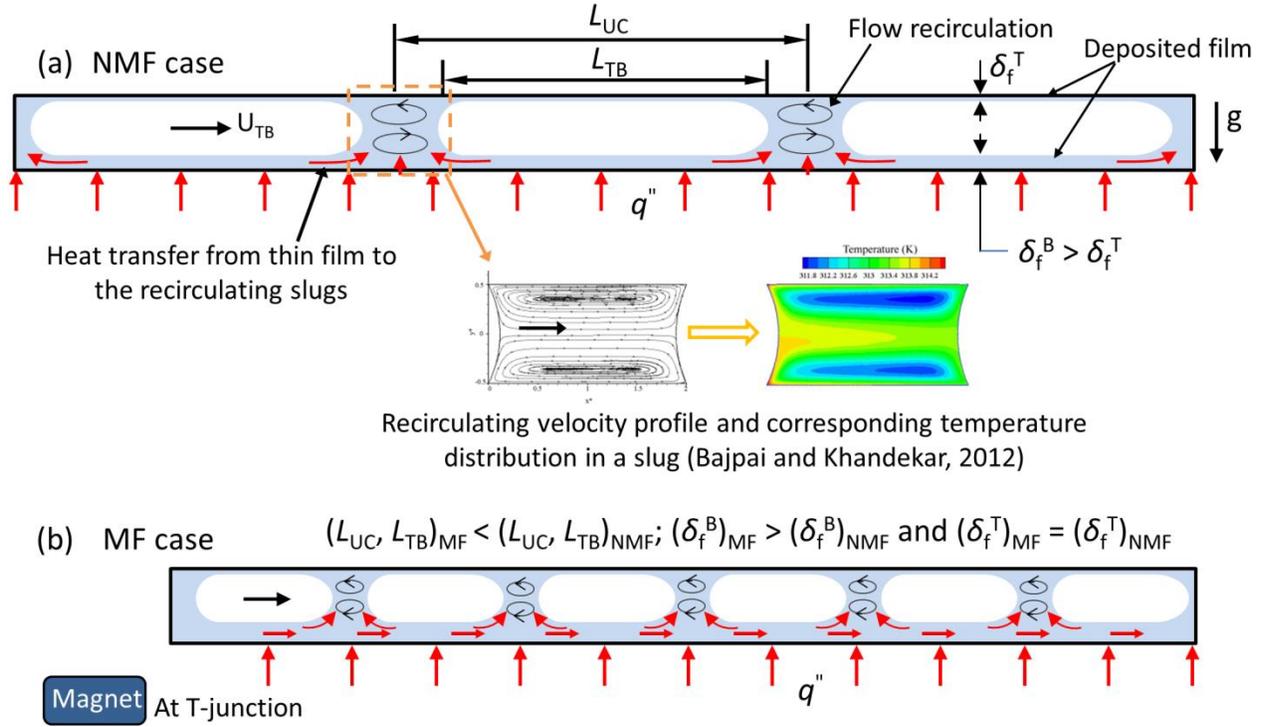

**Figure 9:** Schematics of the heat transfer models in (a) NMF (b) MF cases of the TBF.

### 4.2.1 Effect of heater wall orientation

As explained earlier, the distribution of liquid film across the bubbles is asymmetric because of the buoyancy effect. To investigate the effect of film thickness on the heat transfer in both NMF and MF cases, the orientation of the experimental setup is changed to make the heated wall work as the bottom, top and sidewall of the channel in different experiments. A comparison of plots of $\bar{T}_{wall}$ and $h_{local}$ for all three cases of heater wall orientations (bottom, top, and side) for NMF and MF ($B = 0.19$ T) cases are presented in Figures 10 and 11 for FF #2. Another comparison of $Nu_{avg}$ and net change, $\eta = (((Nu_{avg})_{MF} - (Nu_{avg})_{NMF})/(Nu_{avg})_{NMF})$ due to MF is presented in Figure 12 for both FF #1 and FF #2 ferrofluids, for all three configurations.

As seen from the plots of $\bar{T}_{wall}$, the decrease in the wall temperatures is lowest for the bottom wall heating case than side and top wall cases. The gas fraction is also high ($\beta = 0.70 - 0.80$) in those cases. As liquid and gas phases approach the same flow rates ($\beta \leq 0.50$), the size of generated Taylor bubbles gets smaller even in NMF cases compared to higher $\beta$ cases, as seen in Figure 5. The effect of magnetic manipulation diminishes as the size of Taylor bubbles become smaller in NMF cases. This also reflects on the local and average heat transfer, as seen in the plots of $h_{local}$ and $Nu_{avg}$. The enhancement in heat transfer is highest for the bottom wall heating case at high gas fractions, as seen from the plots in Figures 11 and 12. As *β* approaches 0.50, the enhancement factor diminishes (*η* ~ 15% - 30%) for all the cases.

It is also seen from the plots that for the top wall heating case, the generation of bubbly flow through magnetic manipulation is not beneficial compared to side and bottom wall cases. This is due to the fact that the liquid film thickness and the interaction of the liquid phase are at their minimum in the top wall heating case due to the effect of gravity, as compared to the other two cases. The generation of bubbly flow through magnetic manipulation at high gas fractions does not improve the interaction of the liquid phase with the heated wall like in other cases. Therefore, the heat transfer is nearly identical in both NMF and MF cases in top wall heating as liquid phase interaction with the heated wall is approximately the same in both cases. This is unlike the bottom wall heating case, where the interaction of the liquid phase increases substantially with such flow modifications. The augmentation in the sidewall heating cases is lower than the bottom wall case but higher than the top wall heating case. The interaction of the liquid phase with the heated surface and the thickness of the liquid film is expected to lie between the lower limits of the top wall case and the upper limits of the bottom wall heating case.

### (a) Heater as Bottom wall

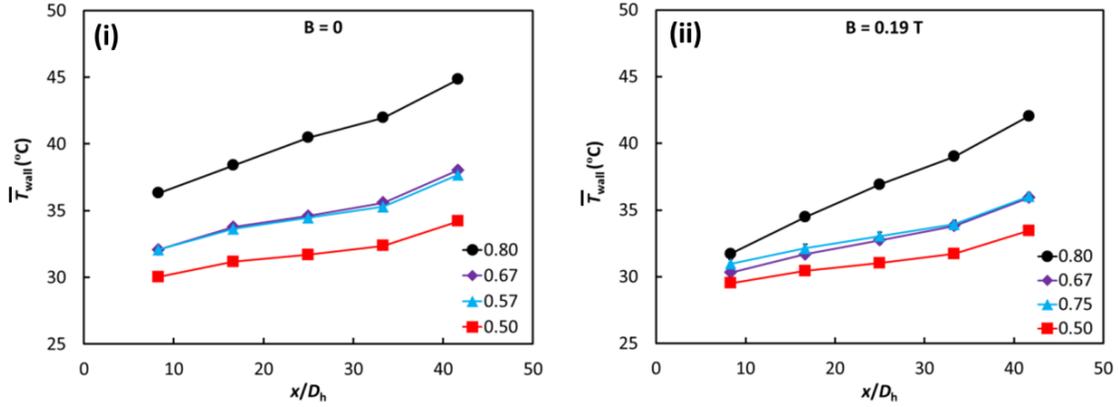

### (b) Heater as Top wall

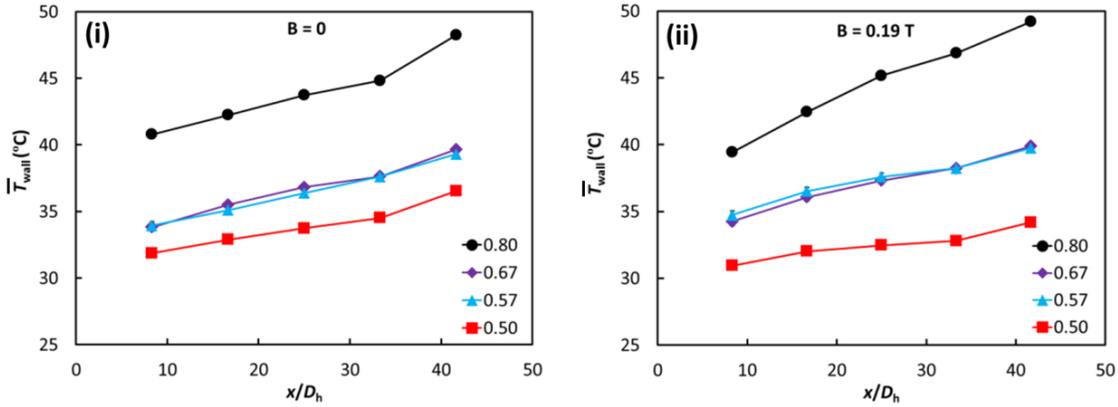

### (c) Heater as Side wall

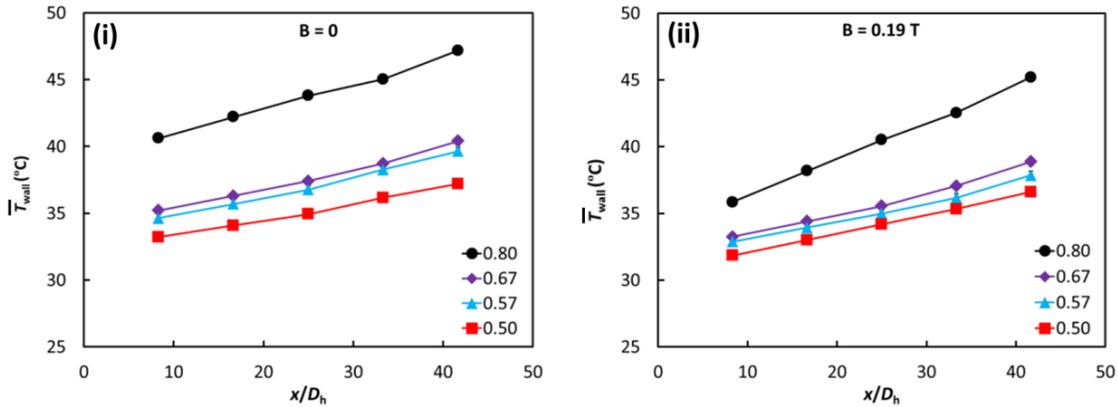

**Figure 10:** Comparison of plots of $\overline{T}_{wall}$ at (i) $B = 0$ (NMF) and (ii) $B = 0.19$ T (MF) for (a) Bottom (b) Top (c) Side wall heater configuration for FF #2 at different β.

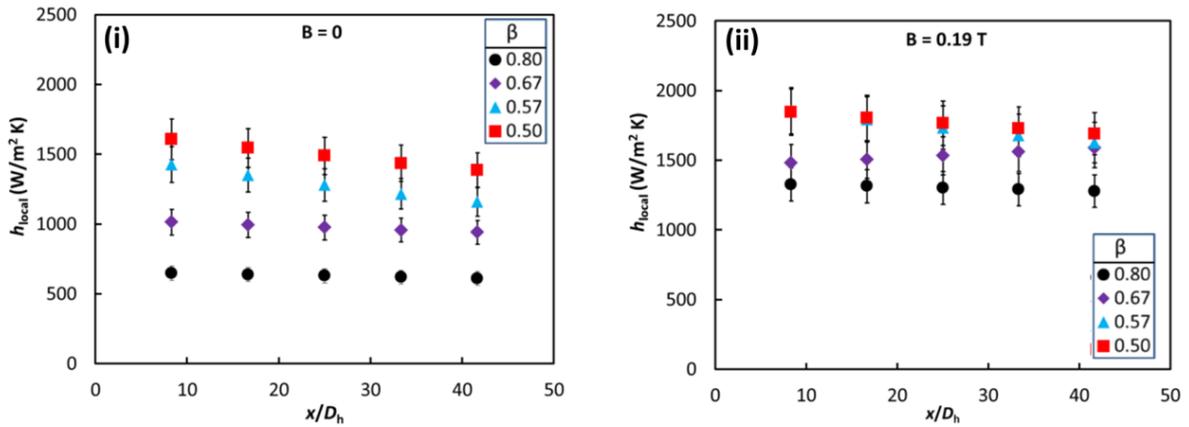

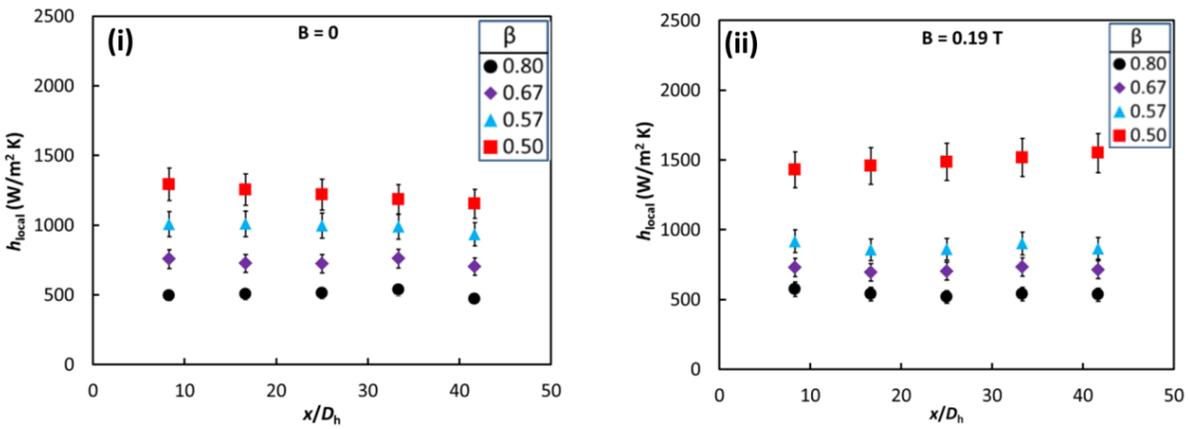

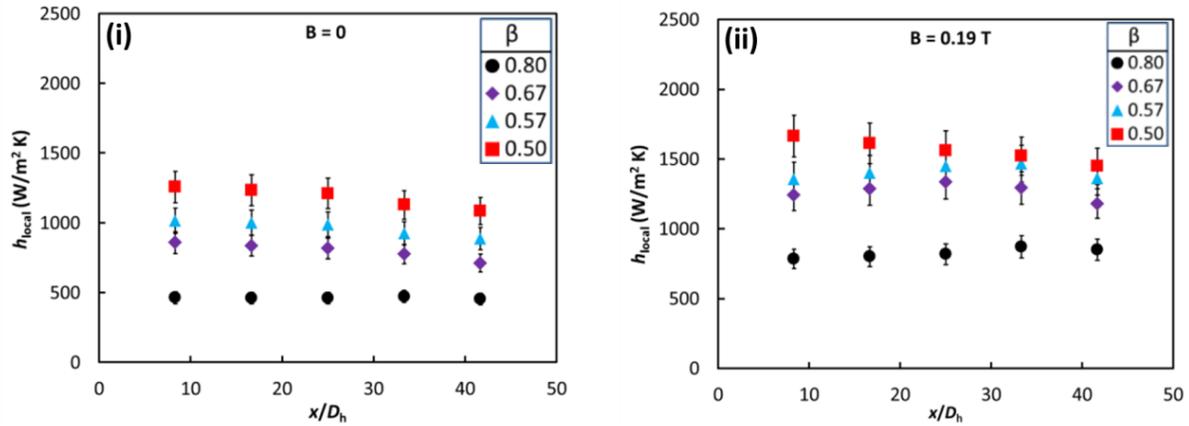

**Figure 11:** Comparison of plots of $h_{local}$ for (a) Bottom (b) Top (c) Side wall heater configuration for FF #2 at (i) $B = 0$ (NMF) and (ii) $B = 0.19$ T (MF).

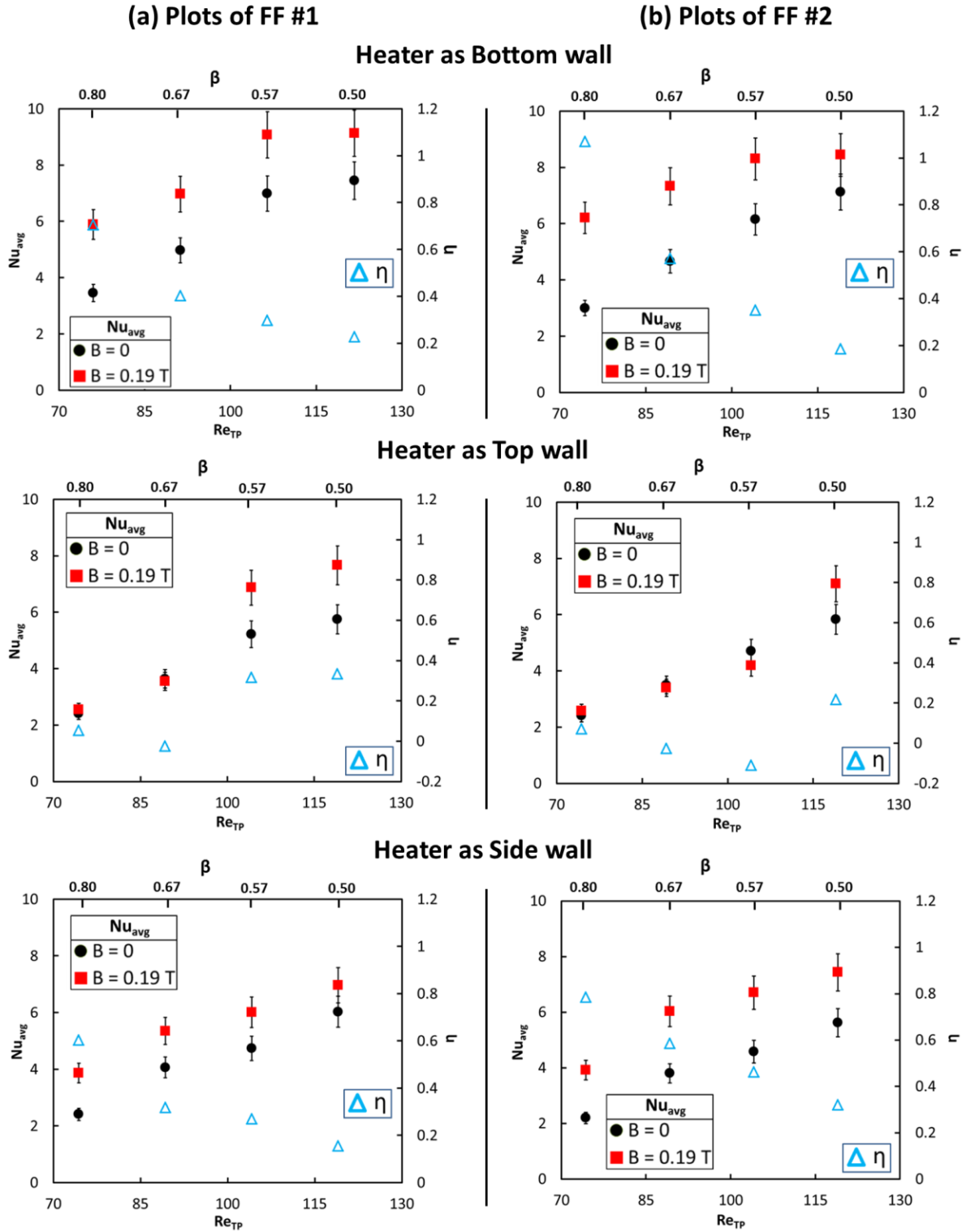

**Figure 12:** Comparison of plots of $Nu_{avg}$ and net change ($\eta$) against the $Re_{TP}$ and $\beta$ for (a) FF #1 (b) FF #2 ferrofluids in (a(i), b(i)) Bottom (a(ii), b(ii)) Top (a(iii), b(iii)) Side wall heating cases for $B$ = 0 (NMF) and $B$ = 0.19 T (MF).

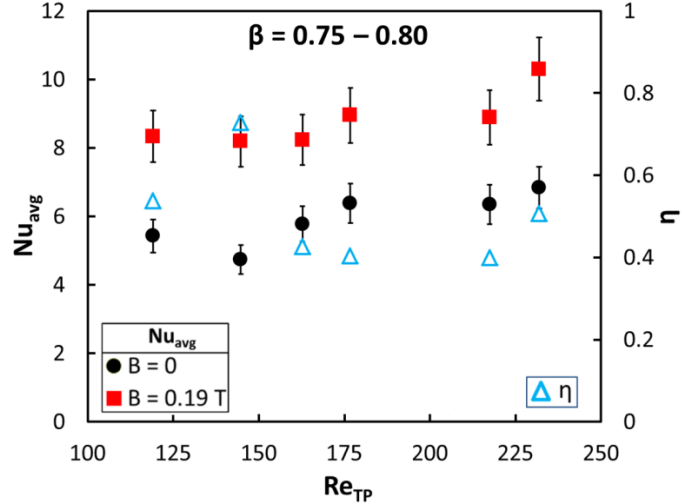

**Figure 13:** Plots of $Nu_{avg}$ and $\eta$ against the $Re_{TP}$ for FF #2 ferrofluids in bottom wall heating configuration for $B$ = 0 (NMF) and $B$ = 0.19 T (MF) cases when β is kept in the range 0.75 - 0.80.

It is evident from the observations that the augmentation is highest for the bottom wall heating case at high gas fractions (*β* ~ 0.70 - 0.80). To further examine this trend, the investigation of the bottom wall heating case was extended by changing the flow rates of air and FF #2 ferrofluid in the range $Q_{air}$ = 15 - 30 ml/min and $Q_{ff}$ = 5 - 10 ml/min, respectively, to keep *β* in the range 0.75 - 0.80, and the effect of magnetic manipulation is examined in all such cases. Figure 13 shows the plot of $Nu_{avg}$ and $\eta$ against $Re_{TP}$ for this extended study.

As seen from the plots, the net gain is in the range 40% - 60% for the studied range of $Re_{TP}$ against the highest enhancement of 100% and 70% observed at the lowest studied flow rates of $Q_{ff}$ = 2.5 ml/min and $Q_{air}$ = 10 ml/min for the two ferrofluids. The observed decrease in the net enhancement can be attributed to the fact that the increase in the flow rate (or Ca) of ferrofluids, the mechanism of bubble formation gets affected, and relatively smaller lengths of bubbles/unit-cells are formed at higher flow rates, even when gas fractions (*β*) is kept same. That means more numbers of relatively smaller bubbles and unit-cells will generate at higher flow rates. This can also be verified from Figure 13, in which $Nu_{avg}$ increases as $Re_{TP}$ is increased in NMF cases. In the preceding section, it was seen that the effect of magnetic manipulation gets weaker as the length of bubbles decreases. It can also be verified from the investigations of Leung et al. [17, 19], in which different lengths of bubbles and liquid slugs were generated at the same mixture velocity and gas fractions by using different sizes of T-junctions. Overall, the net enhancement stays around ~ 50% for the extended study, and it is expected to continue this trend as long as the length scale of bubbles is comparable to the heater.

## 5. Summary and Conclusions

The thermal transport characteristic of the magnetically manipulated Taylor bubble flow of ferrofluids is investigated in the present study. A permanent magnet is used to create a spatially distributed magnetic field locally. The applied magnetic field induces a magnetic force in ferrofluids, which alters the mechanism of Taylor bubble formation when placed in the vicinity of the air-ferrofluid interface forming at the T-junction. The strength of the induced magnetic force is varied to change the size of generated bubbles at the given flow rates. Through magnetic manipulation, smaller sizes ($L_{TB} < 3D_h$) of Taylor bubbles and unit-cells are generated in the resulting two-phase flow, while flow rates of the respective phases are kept unchanged. The generation of smaller bubbles/unit-cells causes substantial modifications in the resulting two-phase flow: (i) The time-averaged void fraction of the resulting two-phase flow decreases (ii) Higher number of units interact with the heater at any time instance (iii) Frequent disturbances in the boundary and interfacial regions of the flow are created compared to larger bubble-slug system. These flow modifications significantly augment the heat transfer in the resulting two-phase flow. The extent of augmentation is observed to be dependent on multiple parameters, which are examined in the present work. Following major conclusions are drawn from the present study:

- The application of ferrofluids in Taylor bubble/slug flow with magnetic field-assisted alteration of the flow morphology can be used for achieving an on-demand augmentation in thermal transport characteristics (which can be up to 100%).
- Augmentation through magnetic manipulation is more effective in high gas volume fractions (~ 0.70 – 0.80) cases, in which generated bubbles are large and comparable with the length scale of the heater. The net augmentation can also be tuned in between an upper and a lower limit by varying the induced magnetic force. The magnetic force can be changed either by changing the volume fraction of the nanoparticles in the bulk liquid or the gradient of the applied magnetic field.
- The void fraction of the gas phase, or alternatively the liquid film thickness around the Taylor gas bubble of the resulting two-phase flow plays a dominant role in overall heat transfer enhancement. Magnetic-field assisted alteration of Taylor bubble flows had a lower void fraction (hence, higher thickness of the liquid film), which contained an additional supplementary volume of liquid participating in the heat exchange process compared to the flow with larger bubbles ($L_{TB} > 3D_h$) and, therefore, exhibiting higher heat transfer.

- The generation of highly bubbly flow through magnetic manipulation did not improve heat transfer in the top wall heating cases, as compared to the bottom and side wall heating cases at high gas fractions (> 0.50). This is because the interaction of the liquid phase with the heated surface is at the lowest for the top wall case due to flow drainage under the effect of gravity. These observations also indicate that the liquid phase is the main contributor to the overall heat exchange process with flow circulation caused by the Taylor bubbles, and sufficient interaction of the liquid phase with the heated wall is required for any meaningful augmentation through magnetic manipulation. As the flow rate of ferrofluids is increased equal to the gas flow rate, a higher amount of liquid comes into contact with the heated top surface, and then magnetic manipulation shows marginal enhancement (~ 15% - 30%) in the heat transfer, similar to other cases.

The proposed application of ferrofluids with the suggested magnetic manipulation technique provides a useful alternative for an on-demand augmentation in two-phase heat transfer for low Reynolds number flows. To further enhance the heat transfer, an immiscible liquid can be used as the dispersed phase in place of gas/air. Such strategies will be beneficial for small-scale systems such as microfluidics and mini/micro-scale two-phase heat exchanges for which flow is mostly in the laminar regime with inferior transport characteristics. Such systems can also be used to study the effect of phase distribution, void fraction, and liquid film thickness on thermal transport characteristics of two-phase flows while geometrical and flow parameters are identical.

## Acknowledgement

The authors would like to thank Dr. Madhushree Kole, Postdoctoral associate at the Department of Mechanical Engineering, IIT Kanpur, for her useful suggestions. The authors would also like to thank Dr. Kinnari Parekh, Charotar University of Science and Technology (CHARUSAT), Gujrat (India), for supplying the required ferrofluid. The authors are grateful to IIT Kanpur for providing resources and facilities to conduct this study. This research did not receive any specific grant from funding agencies in the public, commercial, or not-for-profit sectors.

## Appendix:

## Uncertainty analysis

Uncertainties in the measurements and computations arise from different sources for the measured quantities. Single sample uncertainty analysis proposed by Kline and McClintock (1953), is applied to data analysis in this work. Equation A1 shows the method for the estimation of uncertainty in a quantity $X$ which depends on other quantities $x_1, x_2 \ldots \ldots x_n$ as follows,

$$\delta X = \left[\left(\frac{dX}{dx_1}\delta x_1\right)^2 + \left(\frac{dX}{dx_2}\delta x_2\right)^2 + \cdots + \left(\frac{dX}{dx_n}\delta x_n\right)^2\right]^{1/2} \qquad \text{A1}$$

The uncertainty in the measurements and computations of different physical quantities is estimated as follows,

Uncertainty in flow rate/superficial velocity comes from the accuracy in the dispensing of syringe pumps. For the measurement of Taylor bubble/unit-cell lengths and their normalized values, uncertainty comes from defining pixel resolution of the known distance and the measurement of unit-cell length in pixels from the acquired images, as follows,

$$\frac{\delta(L_{TB}^* \text{ or } L_{UC}^*)}{(L_{TB}^* \text{ or } L_{UC}^*)} = \left\{\left(\frac{\delta(L_{TB} \text{ or } L_{UC})}{(L_{TB} \text{ or } L_{UC})}\right)^2 + \left(\frac{\delta D_h}{D_h}\right)^2\right\}^{1/2} \qquad \text{A2}$$

Heat flux is computed as, q" = (V × I)/Area = V × I/(L ×W). Where V and I are voltage and current. L and W are the length and width of the SS heater strip. Uncertainty in heat flux is estimated as follows,

$$\frac{\delta q''}{q} = \left\{\left(\frac{\delta V}{V}\right)^2 + \left(\frac{\delta I}{I}\right)^2 + \left(\frac{\delta L}{L}\right)^2 + \left(\frac{\delta W}{W}\right)^2\right\}^{1/2} \qquad \text{A3}$$

The heat transfer coefficient is computed as $h = q''/(\bar{T}_{\text{wall}} - \bar{T}_{\text{fluid}})$. Uncertainty in $h$ is estimated as follows,

$$\frac{\delta h}{h} = \left\{\left(\frac{\delta q''}{q}\right)^2 + \left(\frac{\delta \Delta T}{\Delta T}\right)^2\right\}^{1/2} \quad \text{A4}$$

$$\delta(\Delta T) = \left\{(\delta T_{wall})^2 - (\delta T_{fluid})^2\right\}^{1/2}$$

Table A1 shows the estimation of maximum possible uncertainties in the measured and computed quantities.

Table A1: Maximum uncertainty in the measured/computed variables

| S. no. | Quantity | Range | Maximum uncertainty |
|---|---|---|---|
| 1. | Flow rate ($Q$) | 2.5 - 30 ml/min | ±1.0% |
| 2. | Temperature ($T$) | 20 - 50°C | ±0.3°C |
| 3. | Voltage ($V$) | 0 - 60 V | ±0.5% |
| 4. | Current ($I$) | 0 - 6 A | ±0.5% |
| 5. | Image resolution | 40 pixels/mm | ±2.7% |
| 6. | $L_{TB}^*$, $L_{UC}^*$ | 0.50 - 40 | ±7.4% |
| 7. | Heat flux ($q''$) | 7 - 10 kW/m² | ±8.9% |
| 8. | Heat transfer coefficient ($h$) | 500 – 2000 (W/m²K) | ±9.0% |
| 9. | Nusselt number (Nu) | 2 - 10 | ±9.0% |

## Thermocouple calibration

Thermocouples were calibrated using a temperature bath (Thermo-Haake® DC10-K20; ±0.1°C, with RTD PT-1000 NIST traceable calibration) and found to be within the specified uncertainty range (± 0.3°C). Sample calibration curves of two thermocouples are provided below.

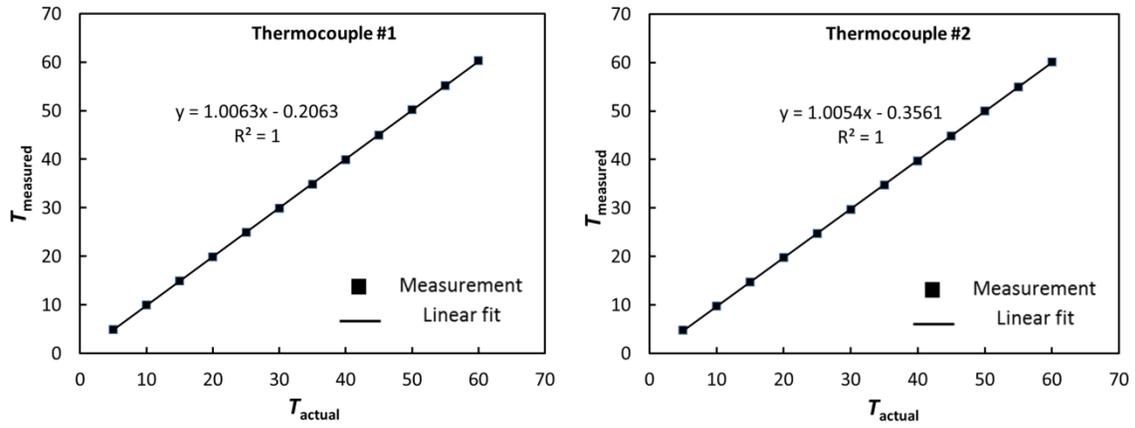

Figure A1: Sample calibration curves of thermocouples

S. Kline and F. Mcclintock, 1953 "Describing Uncertainties in Single-Sample Experiments," Mechanical Engineering, Vol. 75, pp. 3-8.